\documentclass{SciPost}

\hypersetup{
    colorlinks,
    linkcolor={red!50!black},
    citecolor={blue!50!black},
    urlcolor={blue!80!black}
}

\usepackage[bitstream-charter]{mathdesign}
\usepackage{comment}
\DeclareSymbolFont{usualmathcal}{OMS}{cmsy}{m}{n}
\DeclareSymbolFontAlphabet{\mathcal}{usualmathcal}

\fancypagestyle{SPstyle}{
\fancyhf{}
\lhead{\colorbox{scipostblue}{\bf \color{white} ~SciPost Physics }}
\rhead{{\bf \color{scipostdeepblue} ~Submission }}

\fancyfoot[C]{\textbf{\thepage}}
}

\UseRawInputEncoding

\begin{document}

\pagestyle{SPstyle}

\begin{center}{\Large \textbf{\color{scipostdeepblue}{
Quantum Ising transition on the Cayley tree:\\boundary criticality, universality, and an effective Bethe line
}}}\end{center}

\begin{center}\textbf{
Pietro Richelli\textsuperscript{1$\star$} and
Natalia Chepiga\textsuperscript{2$\dagger$} 
}\end{center}

\begin{center}
{\bf 1} Kavli Institute of Nanoscience, Delft University of Technology,\\
Lorentzweg 1, 2628 CJ Delft, the Netherlands \\
{\bf 2} The Rudolf Peierls Centre for Theoretical Physics, Oxford
  University, Oxford OX1 3PU, United Kingdom
  
$^\star$\href{mailto:p.richelli@tudelft.nl}{\small p.richelli@tudelft.nl}\,,
$^\dagger$\href{mailto:N.Chepiga@tudelft.nl}{\small natalia.chepiga@physics.ox.ac.uk}
\end{center}

\section*{\color{scipostdeepblue}{Abstract}}
\textbf{\boldmath{
We investigate quantum criticality in the transverse-field Ising model on finite Cayley trees and demonstrate that their extensive boundary fundamentally changes the nature of the quantum phase transition compared with the infinite, boundary-free Bethe lattice. Using tree tensor-network simulations for systems with up to $N=12286$ sites, we identify a boundary-driven continuous transition at $h_x^c\approx1.76$, distinct from the mean-field Bethe-lattice transition at $h_x^c\approx2.23$. We characterize this transition through finite-size scaling of the magnetization and magnetic susceptibility and extract a consistent set of critical exponents. By introducing next-nearest-neighbor interactions, we demonstrate the robustness  of the extracted critical exponents, providing numerical argument for the Ising boundary universality class on the Cayley tree. Furthermore, we show that the physics of the boundary-free Bethe lattice can be efficiently recovered on a finite Cayley tree by fine-tuning a local boundary field. The resulting \textit{effective Bethe line}, whose existence is tied to the mean-field nature of the bulk transition, effectively screens the extensive boundary and restores both the critical point and the mean-field critical scaling of the infinite Bethe lattice.
}}

\vspace{\baselineskip}



\vspace{10pt}
\noindent\rule{\textwidth}{1pt}
\tableofcontents
\noindent\rule{\textwidth}{1pt}
\vspace{10pt}

\section{Introduction}
\label{sec:intro}

The study of quantum phase transitions in low-dimensional lattice models is one of the most active areas of research in condensed matter physics\cite{Sachdev_2011}. Low-dimensional systems exhibit enhanced quantum fluctuations, making them a fruitful ground for exotic phases and nontrivial critical behavior. Quantum critical phenomena are traditionally associated with continuous phase transitions characterized by the spontaneous breaking of a symmetry in the bulk. However, as discussed by Cardy \cite{cardy2008boundaryconformalfieldtheory,CARDY1986200}, their manifestation in finite-size systems can be significantly modified by boundary critical phenomena. This has motivated the development of boundary conformal field theory (bCFT) and the extension of many established universality classes, primarily formulated for conformally invariant transitions, to incorporate nontrivial boundary critical behavior\cite{Affleck_potts,ashkintellechepiga,chepiga_potts,Ian_Affleck_2000}.   
In these settings, however, the boundary typically produces only sub-leading modifications, leaving both the nature and the location of the bulk transition unchanged. For instance, boundary conditions can select different primary fields within the same underlying critical field theory or control sub-leading corrections to otherwise universal critical scaling. 
In this paper, we will show that on tree lattices the boundary plays a decisive role, affecting both the nature and the location of the quantum phase transition. This gives rise to two fundamentally distinct types of quantum phase transitions on trees: bulk and boundary quantum phase transitions.
To date, most studies of quantum critical phenomena on tree lattices have focused primarily on the Bethe lattice an infinite, symmetric, regular, and cycle-free tree \cite{bethe_lattice}. Classical models on the Bethe lattice have been studied extensively, with many admitting exact analytical solutions \cite{spectrum_cayley_bethe,axioms11120703}.
More recently, a range of quantum models have been investigated using tensor-network methods, including the transverse-field Ising model \cite{TFIM_Nagaj,Li_efficient,daniskaAnalysisQuantumSpin2016}, the Bose-Hubbard model \cite{Bose_Hubbard_bethe}, the antiferromagnetic XXZ model \cite{XXZ_bethe,schmidt_gap_XXZ}, and quantum spin-glass models \cite{mossiQuantumSpinGlass2017}, as well as  models of unitary quantum many-body dynamics \cite{quantum_circuit_bethe}. 
In many of these examples, the quantum critical behavior was found to be of mean-field type, to a certain extent lowering the motivation for further search for non-trivial quantum effects in these systems \cite{Li_efficient}.

\begin{figure}[!ht]
    \centering
    \includegraphics[width=0.4\linewidth]{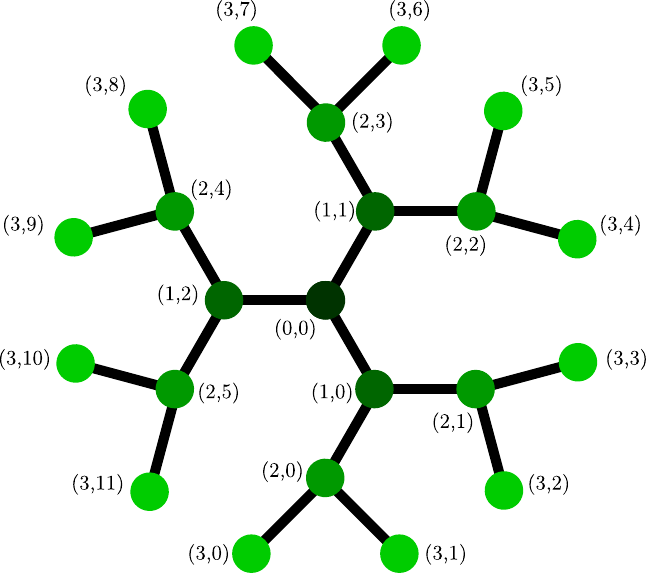}
    \caption{Example of Cayley tree lattice with $n_g =3$ generations. We label lattice sites with two coordinates $(n_i,c_i)$, where $n_i\in \{0,\dots,n_g\}$ marks the generation and $c_i \in \{0,\dots, 3\cdot2^{n_i-1} -1\}$ is a coordinate within each generation. }
    \label{fig:Cayely_tree}
\end{figure}

\noindent
The Cayley tree is the finite-size tree closely related to the Bethe lattice, that exhibit the same bulk structure but, crucially, has a boundary. The size of this finite-size tree is controlled by the number of generations, as sketched in Fig.\ref{fig:Cayely_tree}. The total number of site of a Cayley tree with $n_g$ generations and coordination number $q=3$ is equal $N=3\cdot 2^{n_g}-2$, while the number of boundary sites are $N_b=3\cdot 2^{n_g-1}$. This implies that the number of sites at the boundary always exceed the number of sites in the bulk. In the thermodynamic limit, the boundary remains extensive, with $N_b/N\to 1/2$. Therefore, even in the thermodynamic limit, the Cayley tree does not approach the boundary-free Bethe lattice.
Previous studies on Cayley tree are limited to a few very special models such as the spin $1/2$ Heisenberg model \cite{DensityMatrixRenormalization_XXX}, Rydberg atoms \cite{songQuantumSimulationCayleytree2021}, Ising model with random transverse field \cite{Laflorencie,chakrabartiAbsenceDirectMetal2023,feigelmanSuperconductorinsulatorTransitionEnergy2010}, spin glass \cite{Monthus_2017} and to study Anderson localization \cite{mossi_ergodic_2017}. To the best of our knowledge, the properties of continuous quantum phase transitions in the clean transverse field Ising model and its possible universality remains unexplored and we aim to address this knowledge gap.  
Beyond its fundamental theoretical interest, the study of quantum phenomena on Cayley trees has acquired direct experimental relevance. First of all, the Ising model have already been realized in programmable quantum simulators using neutral atoms on small Cayley trees with up to $N=22$ atoms \cite{PhysRevResearch.3.013286}. It is natural to expect that, in the near future, such experiments will scale to significantly larger system sizes while still operating on finite Cayley-tree geometries with an extensive boundary. Understanding the role of this boundary is therefore essential  to the interpretation of future experiments.
It is worth noting, however, that similar tree structures also arise naturally in synthetic macromolecules known as dendrimers \cite{abbasiDendrimersSynthesisApplications2014} and are of considerable interest in quantum chemistry \cite{nakataniEfficientTreeTensor2013}.

\noindent
The aim of this work is to investigate the nature of the quantum phase transition and its possible universality in the quantum transverse-field Ising model on a Cayley tree.
The rest of the paper is organized as follows. In Sec.~\ref{sec:model_methods} , we define the model and provide an overview of the numerical methods used in this work. In Sec.~\ref{sec:Cayley} , we present the critical scaling of the magnetization, correlations, magnetic susceptibility, and numerically extract the corresponding critical exponents. Furthermore, we investigate the universality of the obtained critical exponents by introducing next-nearest-neighbor interactions \cite{PhysRevResearch.3.013286}.
In Sec.~\ref{sec:EBL} , we demonstrate the existence of an effective Bethe line: by fine-tuning the longitudinal field at the boundary, we show that the finite Cayley tree recovers the critical behavior of the infinite Bethe lattice. Finally, in Sec.~\ref{sec:conclusions}, we summarize our results and put them into a broader perspective.

\section{The model and method}
\label{sec:model_methods}
\subsection{Transverse field Ising model on a tree}
 We study the ground-state properties of the ferromagnetic transverse-field Ising model, defined by the microscopic Hamiltonian:
\begin{equation}
    \mathcal{H}_I = -J\sum_{\langle i,j\rangle} \sigma_i^z \sigma_j^z - h_x \sum_{i} \sigma_i^x\,
    \label{eq:TFIM}
\end{equation}
where $\sigma_i^{x,z}$ denote the Pauli matrices, and the first sum runs over pairs of nearest-neighbor sites on the tree lattice. We label each site $i$ by two coordinates, $(n_i,c_i)$, where $n_i\in{0,\ldots,n_g}$ denotes the generation. Here, $n_i=0$ corresponds to the central site of the tree, while $n_g$ denotes the total number of generations of the Cayley tree, with $n_i=n_g$ corresponding to sites at the boundary. The second coordinate  $c_i \in \{0,\dots, 3\cdot2^{n_i-1} -1\}$ is the coordinate within the given generation. An example of the coordinate assignment for a Cayley tree with three generations is shown in Fig.~\ref{fig:Cayely_tree}. Two sites, $i=(n_i,c_i)$ and $j=(n_j,c_j)$, are nearest neighbors if and only if
$n_i-n_j=1$ and $\left\lfloor \frac{c_i}{2}\right\rfloor=c_j$
where $\lfloor\cdot\rfloor$ denotes the floor function.
The lattice is cycle-free, therefore without loss of generality we set  the coupling parameter $J = 1$ and focus on $h_x>0$. 
Importantly, since the tree lattice is bipartite, the spins on one sublattice can be rotated according to $\sigma^z\rightarrow-\sigma^z$ and $\sigma^y\rightarrow-\sigma^y$, while preserving the spin commutation relations. This transformation reverses the sign of $J$, mapping the ferromagnetic model onto the antiferromagnetic one. We therefore expect the results obtained for the ferro- and antiferromagnetic models to be equivalent, up to an appropriate redefinition of the order parameter and related observables.

\noindent
In one dimension, the transverse field Ising model of Eq.\ref{eq:TFIM} is a textbook example of second order phase transitions separating the ferromagnetic phase with spontaneously broken  $\mathbb{Z}_2$ symmetry and the disordered paramagnetic phase at large $h_x$\cite{Sachdev_2011}. The model is self-dual under Kramers-Wanier duality transformation \cite{kramers_wannier},
exactly solvable \cite{pfeutyOnedimensionalIsingModel1970}, and constitute a lattice realization of the simplest minimal model of conformal field theory (CFT) in 1+1D\cite{difrancescoConformalFieldTheory1997}. It is ubiquitously used as a benchmark in both theoretical and experimental studies and has played an important role in the development of boundary CFT \cite{cardy2008boundaryconformalfieldtheory}.

\noindent
When defined on a tree lattice, the model (\ref{eq:TFIM}) still realizes the two main phases: ferromagnetic with spontaneously broken $\mathbb{Z}_2$ symmetry and a paramagnetic one. However,  beyond one dimension the model is no longer self-dual and the location of the transition is not known exactly. On the (infinite) Bethe lattice the the mean-field transition takes place at $h_x^c \approx 2.23$ \cite{Li_efficient}. However, as we argue below the situation is drastically different on a Cayley tree: we observe a boundary-induced quantum phase transition that  takes place at a different critical point and can be characterized by a different set of critical  exponents.

\subsection{Extended ANNNI model}

We will assess the universality of the transition by studying a paradigmatic anisotropic next-nearest-neighbor Ising model (ANNNI)\cite{ANNNI_1d}. On a tree, the model is defined by the following microscopic Hamiltonian:

\begin{align}
    \mathcal{H}_P = -\sum_{\langle i,j\rangle} \sigma_i^z \sigma_j^z - h_x \sum_{i} \sigma_i^x\ + J_2 \sum_{\langle \langle i,j \rangle \rangle_{n_i \ne n_j}} \sigma_{i}^z \sigma_{j}^z\,, && J_2 >0\,
    \label{eq:ANNNI}
\end{align}
where the notation $\langle \langle i,j \rangle \rangle_{n_i \ne n_j}$ denotes next-nearest-neighbor pairs belonging to different generations. This restriction is introduced to avoid frustration that might lead to spin-glass phases \cite{mossiQuantumSpinGlass2017}.
In contrast to the nearest-neighbor term $J$, the NNN interaction couples sites belonging to the same sublattice, and therefore the sign of $J_2$ cannot be reversed by a local spin rotation. Here, we focus exclusively on antiferromagnetic NNN coupling and restrict ourselves to perturbations with $J_2 \le 0.15$.

\subsection{Tree-DMRG algorithm}

\begin{figure}[!ht]
    \centering
    \includegraphics[width=0.5\linewidth]{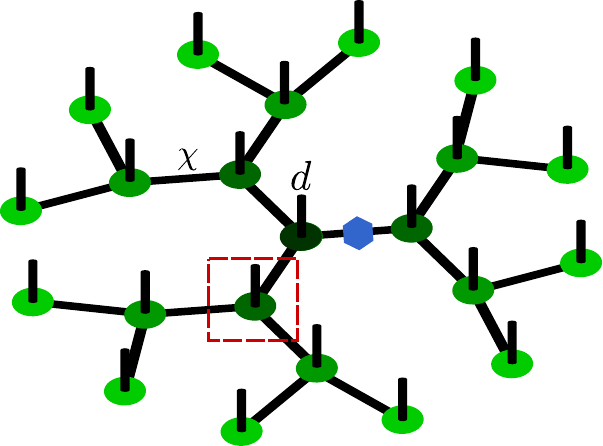}
    \caption{Sketch of a tree tensor network state (TTN) for three generations $n_g = 3$. The blue hexagon represent the singular values in the canonical form. Each tensor has a physical leg (vertical) of size $d=2$ encoding the local Hilbert space of the spin and three  auxiliary legs (or one at the boundary) truncated to the bond dimension $\chi$. Each on-site tensor (green) obeys the normalization condition as discussed in the main text.}
    \label{fig:TTNS}
\end{figure}

We study the ground-state properties of the Ising model on a Cayley tree numerically using a tree version of the DMRG (tDMRG) algorithm\cite{montangero_TTN,Li_efficient}, formulated within the tensor-network framework \cite{ostlund_romer,SCHOLLWOCK201196}. The ground state is represented in terms of local tensors associated with individual lattice sites and connected in a network that directly reflects the geometry of the underlying tree. Consequently, each bulk tensor has four legs, $T^d_{i,j,k}$: one physical leg, encoding the local Hilbert space of the corresponding lattice site, and three auxiliary legs connecting the site to the rest of the tree tensor network. A sketch of the network is presented in Fig.\ref{fig:TTNS}. 

\noindent
On a finite Cayley tree, the boundary tensors have only two legs: a physical leg and a single auxiliary leg, whose bond dimension is bounded by the on-site Hilbert-space dimension $d$ of the boundary site. Moving towards the center of the tree, the Hilbert space associated with a branch grows exponentially with the distance from the boundary. An uncompressed tensor-network representation would therefore require exponentially  increasing bond dimensions. 
For gapped local Hamiltonians, the entanglement entropy is expected to obey an area law, with the relevant area given by the number of bonds connecting the two subsystems. Consequently, for a bipartition obtained by cutting a single bond of the tree, the entanglement remains bounded independently of the exponentially growing number of sites in the corresponding branches. In practice, this allows the auxiliary legs of the tensors to be truncated to a finite bond dimension $\chi$, which, as in standard DMRG, controls the amount of quantum information retained across each bond of the network and provides a systematic parameter for assessing the accuracy of the approximation.
Another important consequence of the fact that a tree can be bipartitioned by cutting a single bond is that one can exploit the gauge freedom of the tensor-network representation to bring the network into a canonical form \cite{comb_TN,ciracMatrixProductStates2021}, in which the local tensors satisfy the following normalization conditions:
\begin{figure}[h!]
    \centering
    \includegraphics[width=.75\textwidth]{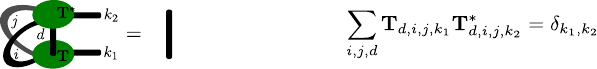}
\end{figure}

\noindent
These canonical conditions allow us to take full advantage of the standard DMRG algorithm by reducing the generalized eigenvalue problem arising from the local tensor optimization to a standard eigenvalue problem \cite{SCHOLLWOCK201196}.

\begin{figure}[!ht]
    \centering
    \includegraphics[width=1\linewidth]{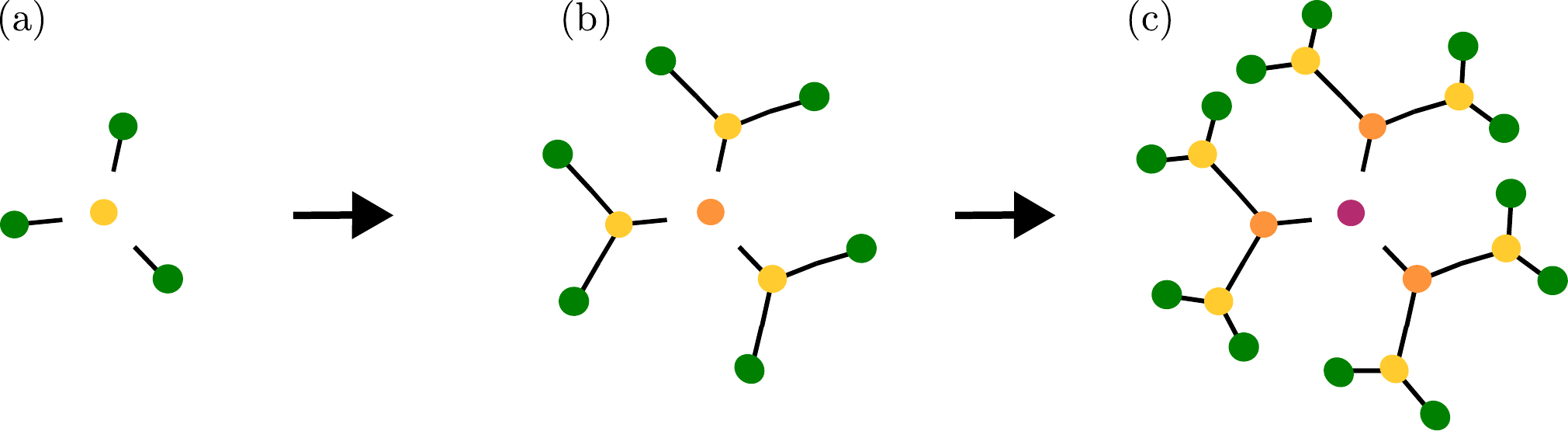}
    \caption{Schematic visualization of the infinite tree-DMRG algorithm used to construct an initial guess for the subsequent finite-size optimization. (a) Starting from identity matrices at the boundary (green), we solve the eigenvalue problem to determine the central tensor (yellow) of a tree with $n_g=1$. (b) We contract copies of the optimized tensor (yellow) with the boundary tensors (green), thereby growing the environment by one generation. We then solve the eigenvalue problem for the enlarged network with $n_g=2$ to determine the new central tensor (orange). (c) We repeat this procedure until the desired system size is reached: at each step, copies of the previously optimized tensor are absorbed into the branches, thereby growing the tree by one generation, and the ground-state problem is solved with respect to the new central tensor (purple). }
    \label{fig:itDMRG}
\end{figure}

\noindent
In addition to the quantum state, we represent the Hamiltonian in Eq.\ref{eq:TFIM} as a projected entangled-pair operator (PEPO), composed of local tensors arranged in the same tree geometry. The PEPO tensors in the bulk and at the central site are denoted by $\mathbf{H}^{i,j}_{k,l,m}$ and $\mathbf{H_c}^{i,j}_{k,l,m}$, respectively, where $i,j$ are the physical indices spanning the local Hilbert space $\mathbb{C}^2$, while $k,l,m\in\{0,1,2\}$ are the auxiliary indices connecting neighboring tensors. The bulk tensor $\mathbf{H}$ is defined as:
\begin{align}
    \mathbf{H}_{0,l,m} = \mathbf{H}_{l,0,m} = \left(\begin{matrix}
        \mathbb{1} & - J\sigma^z & -h_x\sigma^x \\
        \mathbf{0} & \mathbf{0} & \sigma^z \\
        \mathbf{0} & \mathbf{0} & \mathbb{1}
    \end{matrix}\right)\,, &&
    \mathbf{H_c}_{l,0,0} = \mathbf{H_c}_{0,l,0} = \mathbf{H_c}_{0,0,l} = \left( \begin{matrix}
        -h_x \sigma^x \\ 
        \sigma^z \\
        \mathbb{1}
    \end{matrix}\right)\,.
\end{align} 
Following the established DMRG practice, we construct an initial guess for the tensor network by iteratively increasing its size until the tree reaches the desired number of generations $n_g$, in analogy with the infinite-size DMRG algorithm\cite{SCHOLLWOCK201196}. We start from the exact boundary tensors, marked in green in Fig.\ref{fig:itDMRG}(a), which, without loss of generality, are chosen to be identity matrices of size $d\times d$. We then introduce a tensor at the center (yellow), forming a tree with one generation, $n_g=1$, and determine the ground state of this four-site network. A copy of the optimized central tensor is subsequently absorbed into each branch, thereby constructing the environment for a tree with two generations, $n_g=2$, as shown in Fig.\ref{fig:itDMRG}(b). The ground state of the enlarged tree is then obtained by optimizing the new central (orange) tensor. This procedure is repeated iteratively (the next step is shown in Fig.\ref{fig:itDMRG}(c)) until the desired number of generations is reached.

\begin{figure}[!ht]
    \centering
    \includegraphics[width=1\linewidth]{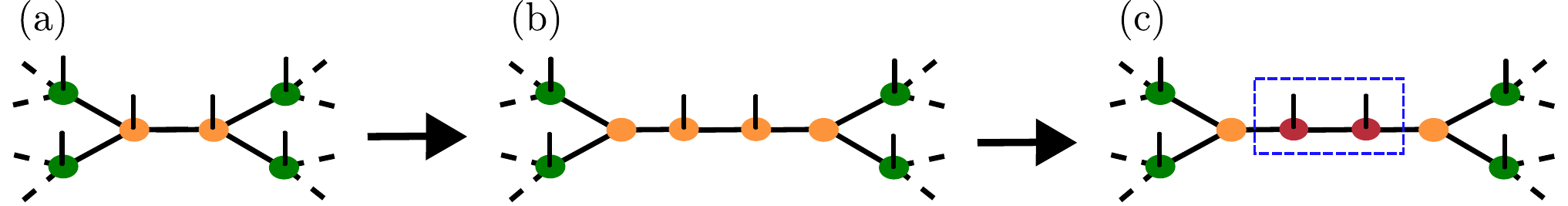}
            \caption{Sketch on the splitting the degrees of freedom to reduce the complexity in the two-site update scheme. (a) Two local tensors that are about to be updated (orange) are connected to the rest of the tree via four bonds. (b) Using SVD decomposition we split each of these rank-4 tensors into two tensors of rank-3. This decomposition is rigorous and does not introduce additional approximation. (c) We update the two tensors with physical bond and truncate the bond dimension connecting them following the standard MPS routine. The new tensors (red) are then recombined with the remainder of the tensors (orange) created in (b).
            }
    \label{fig:2sites}
\end{figure}

\noindent
Once the desired size of the tree is reached, we sweep through the network, optimizing the tensor-network state with respect to the Hamiltonian of Eq.\ref{eq:TFIM}. We employ a two-site optimization scheme: at each iteration, a pair of neighboring tensors is optimized while the rest of the network is kept fixed. Compared to a single-site update, the two-site scheme allows the bond dimension to increase dynamically and is generally more robust against convergence to local minima. However, due to the connectivity of the tree network, a straightforward implementation of the two-site update would be computationally expensive.
To reduce the computational cost, we employ a tensor-splitting procedure commonly used in higher-dimensional tensor-network algorithms, in which the degrees of freedom of each tensor are distributed over multiple tensors. The procedure is sketched in Fig.\ref{fig:2sites} and consists of three steps. {\it (i)} Each of the two tensors to be optimized is decomposed using a singular value decomposition into two parts: a reduced tensor containing the physical leg and the bond connecting the two sites, and a complementary tensor containing the remaining auxiliary legs. No truncation is performed at this stage: all nonzero singular values are retained and absorbed into the reduced tensor participating in the two-site optimization {\it (ii)} The resulting pair of reduced tensors is optimized with respect to the effective two-site Hamiltonian. {\it (iii)} Following the optimization, the tensors are decomposed and recombined with their complementary parts to restore the original structure of the tree tensor network.

\begin{figure}[!ht]
    \centering
    \includegraphics[width=.9\linewidth]{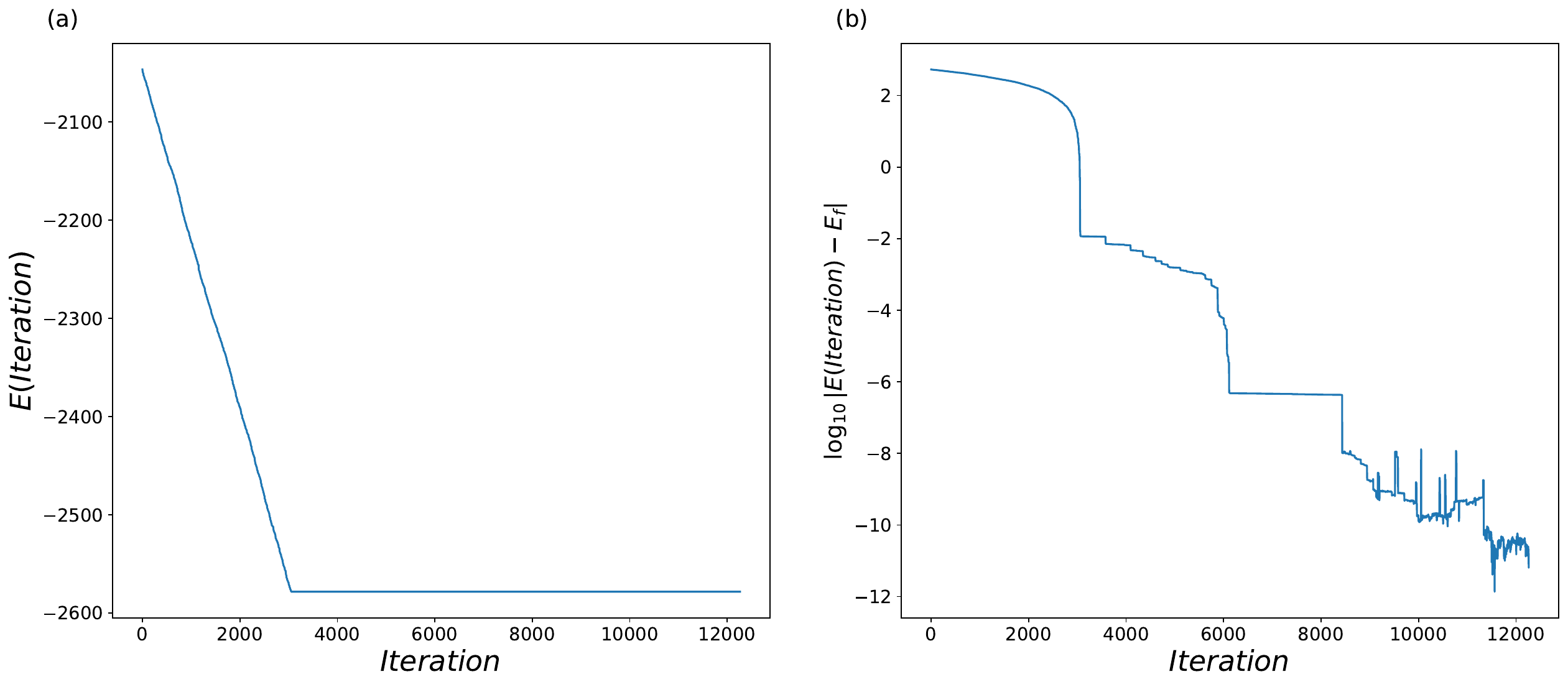}
    \caption{Example of the convergence of the ground-state energy in the transverse field Ising model at $h_x = 1.5$ on a Cayley tree with $n_g = 9$ generations and $N = 1534$ sites, shown on (a) a linear and (b) a semi-logarithmic scale. In (b), since the exact energy is unknown, we plot the difference with respect to the smallest numerical  $E_f$. The fluctuations around $10^{-8}$ indicate the accuracy limit imposed by the chosen truncation error.}
    \label{fig:energyconvergence}
\end{figure}

\noindent
With our algorithm, we are able to reach convergence for trees with up to $n_g = 12$ generations, corresponding to $N = 12286$ sites, with a maximum bond dimension of $\chi=300$. Throughout the entire range of $h_x$ considered in this work, the bond dimension never reaches this maximum value. During the optimization, we discard singular values below $10^{-10}$. We consider the ground state to be converged when its energy changes by less than $10^{-8}$ over a complete sweep through the network. In Fig.\ref{fig:energyconvergence}, we present a representative example of the convergence of the ground-state energy as a function of the number of iterations. 
During each sweep, our algorithm updates the entire tree tensor network state (TTNS), except for the boundary tensors, which, without loss of generality, are kept fixed as identity matrices $\mathbb{1}_{2\times 2}$. Convergence is typically reached within five full sweeps through the network.

\newpage
\section{Cayley tree}
\label{sec:Cayley}

In this section, we investigate the quantum phase transition in the transverse field Ising model defined in Eq.\ref{eq:TFIM} on finite Cayley trees with total generation numbers ranging from $n_g = 6$ to $n_g = 12$. We extract the critical exponent $\beta$, which governs the scaling of the order parameter as the transition is approached from the ordered phase, and establish a lower bound for $\gamma$, which characterizes the divergence of the magnetic susceptibility as the transition is approached from the paramagnetic phase. We also analyze the critical scaling of both quantities at the finite-size transition points. Finally, we examine the behavior of the correlation length across the transition. In the current section we will refer to sites on the boundary for any system size as 

\subsection{Magnetization}
\label{subsec:Cayley_avg_mag}

\begin{figure}[!ht]
    \centering
    \includegraphics[width=1\linewidth]{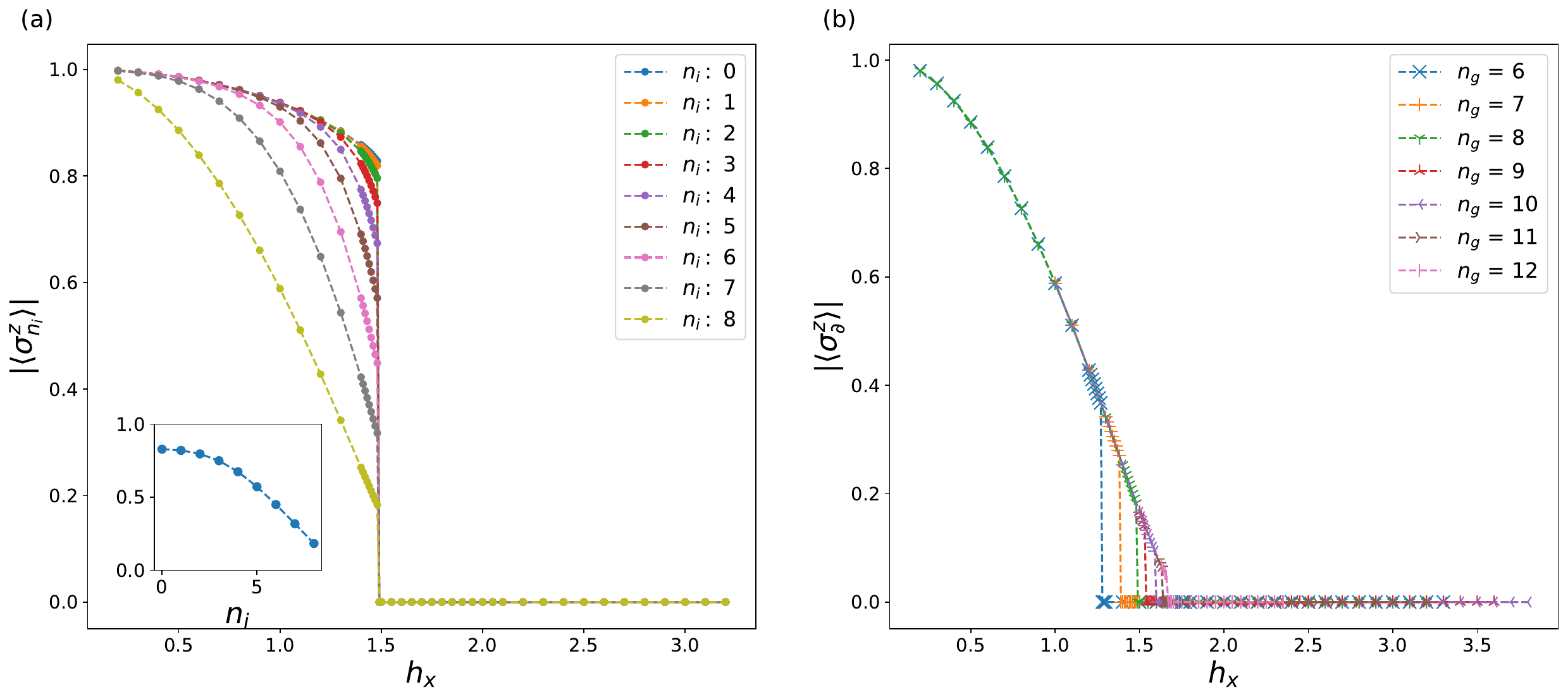}
    \caption{(a) Local magnetization for different generations $n_i$ on a Cayley tree with a total of $n_g = 8$ generations. Here, $n_i=n_g$ corresponds to a boundary site, while $n_i=0$ corresponds to the central spin. The magnetization on each generation exhibits a pronounced jump at the finite-size first-order transition. Note that in the ferromagnetic phase, the boundary magnetization decays much more rapidly than the bulk magnetization. Inset: magnetization just before the finite-size jump as a function of generation index $n_i$.
    (b) Boundary magnetization $|\sigma^z_\partial|$ for Cayley trees with different numbers of generations $n_g$. In the ferromagnetic phase, the boundary magnetization curves for different system sizes overlap perfectly, while the location of the finite-size jump, and hence its height, drifts as the system approaches the thermodynamic limit.  }
    \label{fig:local_mag}
\end{figure}

Many conventional second order quantum phase transitions can be effectively described by Landau-Ginzburg theory\cite{Landau}, which is based on the concept of spontaneous symmetry breaking at the transition. The order parameter is defined to characterize the broken symmetry in the ordered phase and vanishes in the opposite phase, where the symmetry is restored. In the transverse field Ising model, the ferromagnetic phase is traditionally characterized by the spontaneous breaking of the $\mathbb{Z}_2$ spin-flip symmetry, quantified by the on-site magnetization in the bulk or the total magnetization of a finite-size sample. The magnetization is finite in the ferromagnetic phase and vanishes in the paramagnetic phase. The ferromagnetic phase preserves the spatial symmetries of the Cayley tree, and, according to our numerical results, within each generation the local magnetization is homogeneous. However, the most striking feature of the finite-size results is the difference between the scaling of the local magnetization in the bulk and near the boundary of the tree, as shown in Fig.\ref{fig:local_mag}(a).
As mentioned previously, a Cayley tree has more sites at the boundary than in the bulk. For $q=3$, the tree with the slowest growth rate\footnote{Except for the trivial tree with $q=2$ which corresponds to a 1D chain.}, the number of boundary sites is extensive, with $N_\partial - N_{bulk} = 2$. It is therefore not surprising that the boundary sites drive the quantum phase transition in a finite cluster. This feature is particularly evident in Fig.\ref{fig:local_mag}(b), where the boundary magnetization curves not only exhibit the same slope but also overlap perfectly throughout the ferromagnetic phase.

\noindent
The same behavior is observed for sites at a fixed distance from the boundary, as shown in the App.\ref{app:mag_across_tree}. 
Our numerical results indicate that, for any finite-size tree, the transition is first order. However, Fig.\ref{fig:local_mag}(b) clearly shows that the finite-size transition point shifts with increasing tree size, while the discontinuity in the magnetization progressively decreases. This behavior indicates that the transition becomes continuous in the thermodynamic limit, $n_g \to \infty$. Moreover, our results identify the boundary magnetization, rather than the total or bulk magnetization, as the appropriate order parameter for characterizing this transition. Importantly, on the Cayley the transition occurs at $h^c_x\approx1.76$, well before the critical point of the infinite Bethe lattice, $h^c_x\approx2.23$\cite{Li_efficient}.

\subsubsection{Boundary magnetization}
\label{subsubsec:boudnary_mag}
Based on the scaling observed in Fig.\ref{fig:local_mag}(b), we expect the boundary magnetization to vanish continuously as the transition is approached in the thermodynamic limit:
\begin{equation}
    |\langle \sigma^z\rangle|\propto |h_x - h_x^c|^\beta\,,
    \label{eq:mag_critical}
\end{equation}
where $h_x^c$ denotes the critical point in the thermodynamic limit and $\beta$ is the corresponding critical exponent.
In order to locate the critical point more accurately, we analyze the scaling of the boundary magnetization with the distance from the transition.
\begin{figure}[h!]
    \includegraphics[width=1\linewidth]{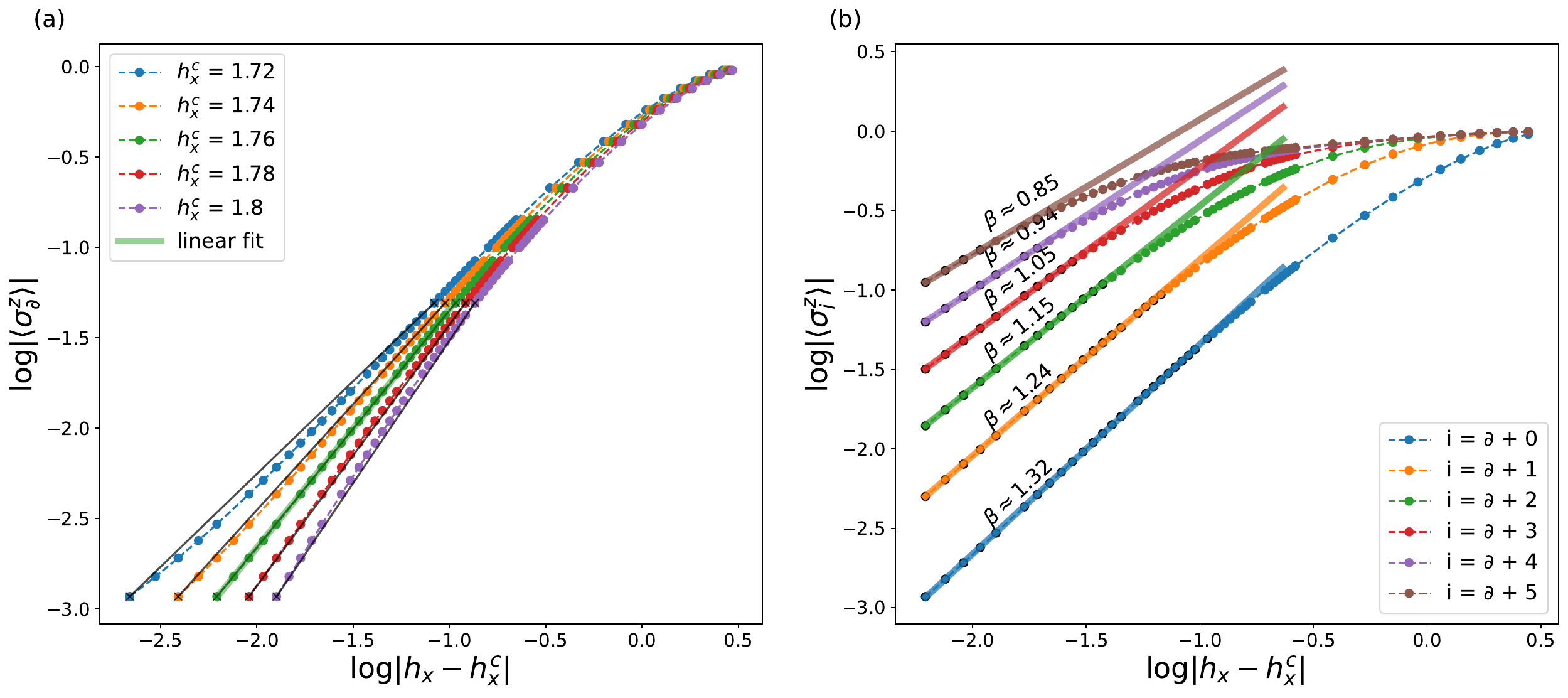}
    \caption{Scaling of the boundary magnetization in the transverse-field Ising model on a Cayley tree. (a) We identify the critical point from the separatrix (green) in the log-log plot of the boundary magnetization versus the distance from the transition, obtained for different candidate values of $h_x^c$. To emphasize the curvature of the different curves, straight lines connecting selected pairs of points are shown as guides to the eye. At larger distances from the transition, the system leaves the critical scaling regime. (b) Scaling of the local magnetization at different distances from the boundary ($\partial + 0$ denotes the boundary sites), assuming $h_x^c \approx 1.76$ as determined in (a). The resulting values of $\beta_i$, indicated above the corresponding slopes, decrease monotonously with increasing distance from the boundary. For each value of $\beta_i$ we highlight the data used for the fit with a black edge.}
    \label{fig:beta_mag}
\end{figure}
Fig.\ref{fig:beta_mag}(a) presents this scaling on a log-log scale for five candidate values of $h^c_x$. We find that for $h_x^c \approx 1.76$ the scaling is essentially linear, forming a separatrix between curves with opposite curvatures obtained for smaller and larger values of $h_x^c$. The critical scaling is visible only sufficiently close to the transition and breaks down once the distance exceeds $|h_x-h^c_x|\gtrsim 0.4$. The slope of the linear regime yields the boundary order-parameter critical exponent $\beta_\partial \approx 1.32$.
In addition to the boundary magnetization, we investigate how the critical scaling evolves upon moving away from the boundary and deeper into the bulk. The corresponding results are presented in Fig.\ref{fig:beta_mag}(b). Although the interval over which the critical scaling can be reliably extracted shrinks noticeably with increasing distance from the boundary, the critical point identified from the boundary magnetization remains consistent with the scaling of the magnetization at other generations: at $h_x^c \approx 1.76$, the corresponding curves remain approximately linear and show no systematic curvature. At the same time, the slope of the scaling, and hence the effective critical exponent $\beta(n_i)$ associated with a given generation, gradually decreases with increasing distance from the boundary. This behavior is consistent with the intuitive picture that, on an infinite tree and infinitely far from the boundary the local properties should approach those of the infinite Bethe lattice, characterized by the mean-field exponent $\beta=1/2$.

\subsubsection{Finite-size scaling of the jump}

\begin{figure}[h!]
    \centering
    \includegraphics[width=1\linewidth]{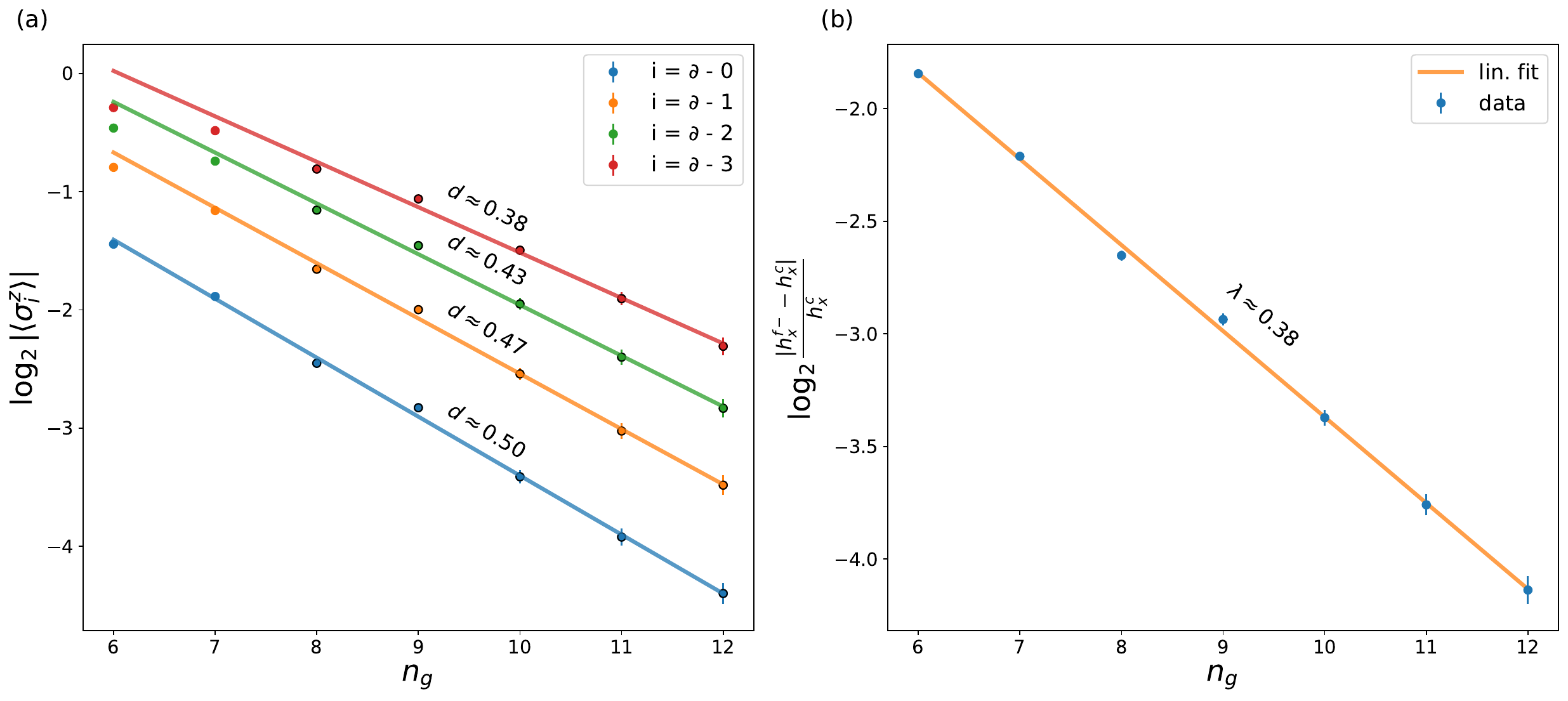}
    \caption{(a) Finite-size scaling of the local magnetization at the boundary (blue) and at distances of up to three generations from the boundary, evaluated immediately before the finite-size jump. In the vicinity of the jump, we vary the field $h_x$ in steps of 0.01; this finite resolution is reflected in the error bars. The magnetization decays exponentially with the number of generations $n_g$ and, equivalently, algebraically with the total number of sites $N$. Lines show linear fits to the data for $n_g\geq8$, highlighted by the black edge. (b) Finite-size scaling of the distance between the location of the magnetization jump and the critical point in the thermodynamic limit as a function of the number of generations $n_g$. The linear fit yields the effective critical exponent $\lambda \approx 0.38$, in the main text we argue that it implies a correlation length critical exponent $\nu=1/\lambda\approx2.63$.}
    \label{fig:bound_mag_scaling}
\end{figure}

The location and the magnitude of the finite-size jumps in the local magnetization shown in Fig.\ref{fig:local_mag}(b) change monotonously with the size of the Cayley tree $n_g$. In this section, we analyze the finite-size scaling of both quantities. More specifically, we measure the local magnetization immediately before the jump at the boundary sites and at distances of up to three lattice cites away from it. The results are presented as a function of the tree size $n_g$ in Fig.\ref{fig:bound_mag_scaling}(a), where the error bars reflect the discrete resolution in $h_x$ used to determine the jumps.
The scaling of the jump magnitude and the finite-size drift of the transition point are not independent. Indeed, as shown in Fig.\ref{fig:bound_mag_scaling}(b), the distance between the finite-size jump and the thermodynamic critical point decreases exponentially with $n_g$ and, equivalently, algebraically with $N$, with an exponent $\lambda \approx 0.38$. We interpret the finite-size jump as occurring when the dominant correlation scale $\xi_\mathrm{dom}$ becomes comparable to the characteristic size of the finite tree. The finite-size drift of the jump therefore provides an indirect probe of the divergence of this correlation scale, which, as we will see in the next section, is otherwise extremely challenging to extract directly on a Cayley tree. Assuming $\xi_\mathrm{dom}(h_x) \propto |h_x-h_x^c|^{-\nu}$ and identifying the finite-size scale with the number of (boundary) sites, $\xi_\mathrm{dom}(h_x^{f-}) \propto N$ we argue that numerically extracted value $\lambda\approx0.38$ corresponds to the critical exponent  $\nu=1/\lambda\approx2.63$.
In other words, our numerical results reveal that the magnitude and the location of the finite-size jump obey the following scaling laws:
\begin{align}
    |\langle \sigma_\partial(h_x^c)\rangle| \propto N^{-d_\partial} \propto 2^{-d_\partial n_g}\,, && \left| \frac{h_x^{f-} - h_x^c}{h_x^c} \right| \propto N^{-1/\nu} \propto 2^{- n_g/\nu}\,.
    \label{eq:scaling}
\end{align}
Substituting Eq.\ref{eq:scaling} into Eq.\ref{eq:mag_critical}, we find that the critical exponents $d_i\,,\beta_i\,,\nu$ are related through the following standard hyperscaling relation:
\begin{equation}
    d_i = \beta_i /\nu\, ,
    \label{eq:hyperscaling}
\end{equation}
where the index $i$ indicates the distance from the boundary, with $i = \partial$ corresponding to the boundary sites.
\begin{table}[h!]
    \centering
    \begin{tabular}{c|c|c|c|c}
        $i =$ & $\partial$ & $\partial+1$& $\partial + 2$ & $\partial +3$ \\ \hline
        $d_i$ & 0.49  & 0.47 & 0.44 & 0.41\\ \hline 
         $\beta_i /\nu$  & 0.50 & 0.47 & 0.43 & 0.40  \\ \hline
           $\beta_i /d_i$  & 2.64 & 2.64 & 2.67 & 2.76 
    \end{tabular}
  
    \caption{Comparison of numerically extracted scaling dimension $d_i$ and the ratio  $\beta_i /\nu$ satisfying the hyperscaling relation in Eq. \ref{eq:hyperscaling} within $3\%$ and the ratio of $\beta_i/d_i$ as an independent measure of the critical exponent $\nu$. The results are in excellent agreement with direct numerical estimate $\nu\approx2.63$ extracted from the scaling of the jump location. }
    \label{tab:hyperscaling}
\end{table}
In Table \ref{tab:hyperscaling}, we compare the scaling dimension $d_i$ extracted from Fig.\ref{fig:bound_mag_scaling}(a) with the ratio of the independently extracted critical exponents $\beta_i/\nu$, obtained from Figs.\ref{fig:beta_mag}(b) and \ref{fig:bound_mag_scaling}(b), respectively. The two estimates agree within a few percent, providing strong numerical support for the scaling relation $d_i=\beta_i/\nu$. Interestingly enough at the boundary, the ratio between $\beta$ and $\nu$ is $1/2$ to a reasonable accuracy. Alternatively, by considering the ratio $\beta_i/d_i$ in the last row of Table \ref{tab:hyperscaling}, we can independently extract the critical exponent $\nu$. The resulting estimates are in excellent agreement with the value obtained from the finite-size scaling of the jump location, $\nu=1/\lambda$, with particularly good agreement at the boundary.

\subsection{Correlations}

In one and two dimensions, the Ising transition is characterized by an algebraic divergence of the correlation length, $\xi \propto |h_x-h_x^c|^{-\nu}$, with $\nu$ being one of the universal critical exponents. In the previous section, we showed how this critical exponent can be extracted indirectly from the finite-size scaling of the magnetization jump. On a tree geometry, however, a direct characterization of the dominant correlations turns out to be particularly challenging. The connected correlations are not determined solely by the graph distance $d$ between two sites, but depend strongly on their positions relative to the boundary. This spatial inhomogeneity becomes increasingly pronounced as the system approaches the transition.
\begin{figure}[h!]
    \centering
    \includegraphics[width=1\linewidth]{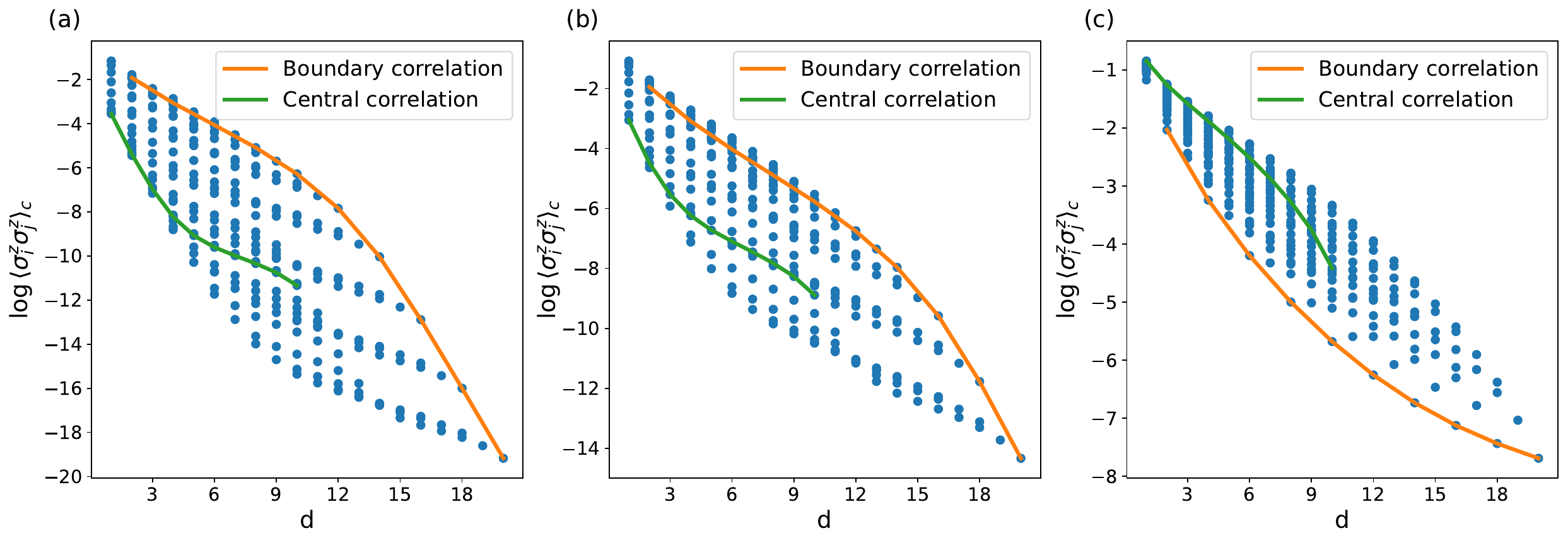}
    \caption{Connected correlation function for a Cayley tree with $n_g = 10$ generations: (a) deep in the ferromagnetic phase at $h_x = 1.5$, (b) immediately before the finite-size jump at $h^*_x = 1.6$, and (c) deep in the paramagnetic phase at $h_x = 1.7$. In all three panels, correlations between pairs of boundary sites and between the central site and one site in each generation are highlighted by the orange and green lines, respectively.}
    \label{fig:CC_cayley}
\end{figure}
In Fig.\ref{fig:CC_cayley}, we present numerical results for the connected correlation function along the longitudinal field $\langle \sigma_i^z \sigma_j^z\rangle_c\equiv \langle \sigma_i^z \sigma_j^z\rangle-\langle \sigma_i^z\rangle\langle \sigma_j^z\rangle$ as a function of the graph distance $d$ between sites $i$ and $j$. We demonstrate that, for a fixed distance $d$, the correlation function can take substantially different values depending on the positions of the two sites relative to the boundary of the Cayley tree. For convenience, correlations between boundary sites are highlighted by the orange line. In Fig.\ref{fig:CC_cayley}(a), we present results deep in the ordered phase; Fig.\ref{fig:CC_cayley}(b) shows the ordered phase close to the finite-size transition point; and Fig.\ref{fig:CC_cayley}(c) shows the system deep in the disordered paramagnetic phase. Remarkably, the boundary correlations, which are among the strongest in the ordered phase, become the weakest in the disordered phase. We argue that this inversion of the spatial structure of correlations is closely connected to the finite-size first-order transition discussed above. 
In Fig.\ref{fig:CC_cayley}, we show all symmetry-inequivalent two-point correlations between sites of the Cayley tree. Interestingly, while the number of sites grows exponentially with $n_g$, the discrete rotational and reflection symmetries of the tree strongly constrain the number of inequivalent correlations, which grows only algebraically with the number of generations. For a Cayley tree with coordination number $q=3$, the total number of inequivalent two-point correlations is
\begin{equation}
    \#\Big(\langle \cdot , \cdot\rangle\Big) = \frac{n_g(n_g+1)(n_g+5)}{6} \propto n_g^3\,.
\end{equation}

\subsection{Magnetic susceptibility}

\begin{figure}[h!]
    \centering
    \includegraphics[width=1\linewidth]{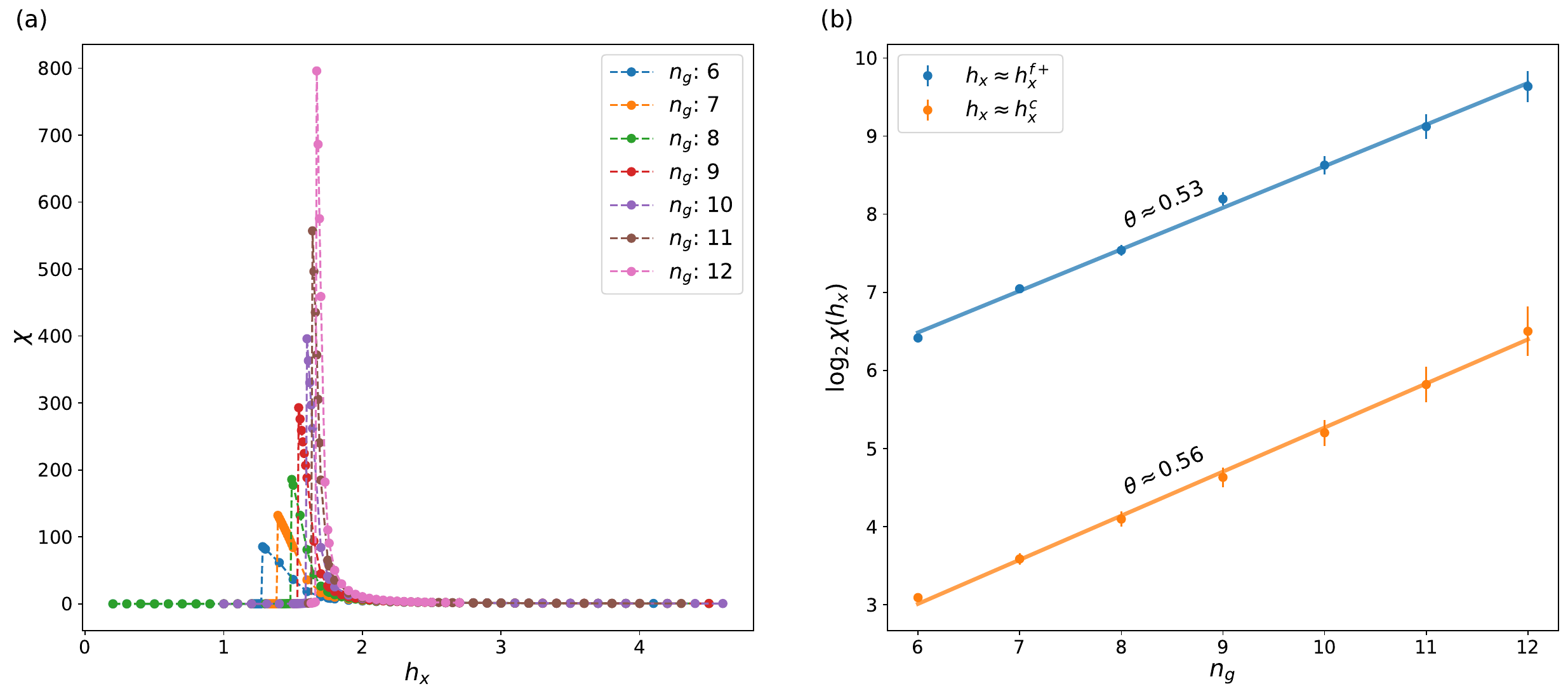}
    \caption{Magnetic susceptibility of the central site on the Cayley tree. (a) Magnetic susceptibility $\chi$ as a function of $h_x$ across the transition for different system sizes, showing the characteristic finite-size jumps. (b) Finite-size scaling of the susceptibility evaluated immediately after the jump, $\chi(h_x^{f+})$ (blue) and at the transition point $\chi(h_x^{c})$ (orange). The extracted scaling exponent is in the range $\theta\approx0.53-0.56$, identified with $\theta=\gamma/\nu$, yields $\gamma=\theta\cdot\nu\approx1.39-1.47$.}
    \label{fig:mag_susc_cayley}
\end{figure}

\begin{figure}[h!]
    \centering
    \includegraphics[width=1\linewidth]{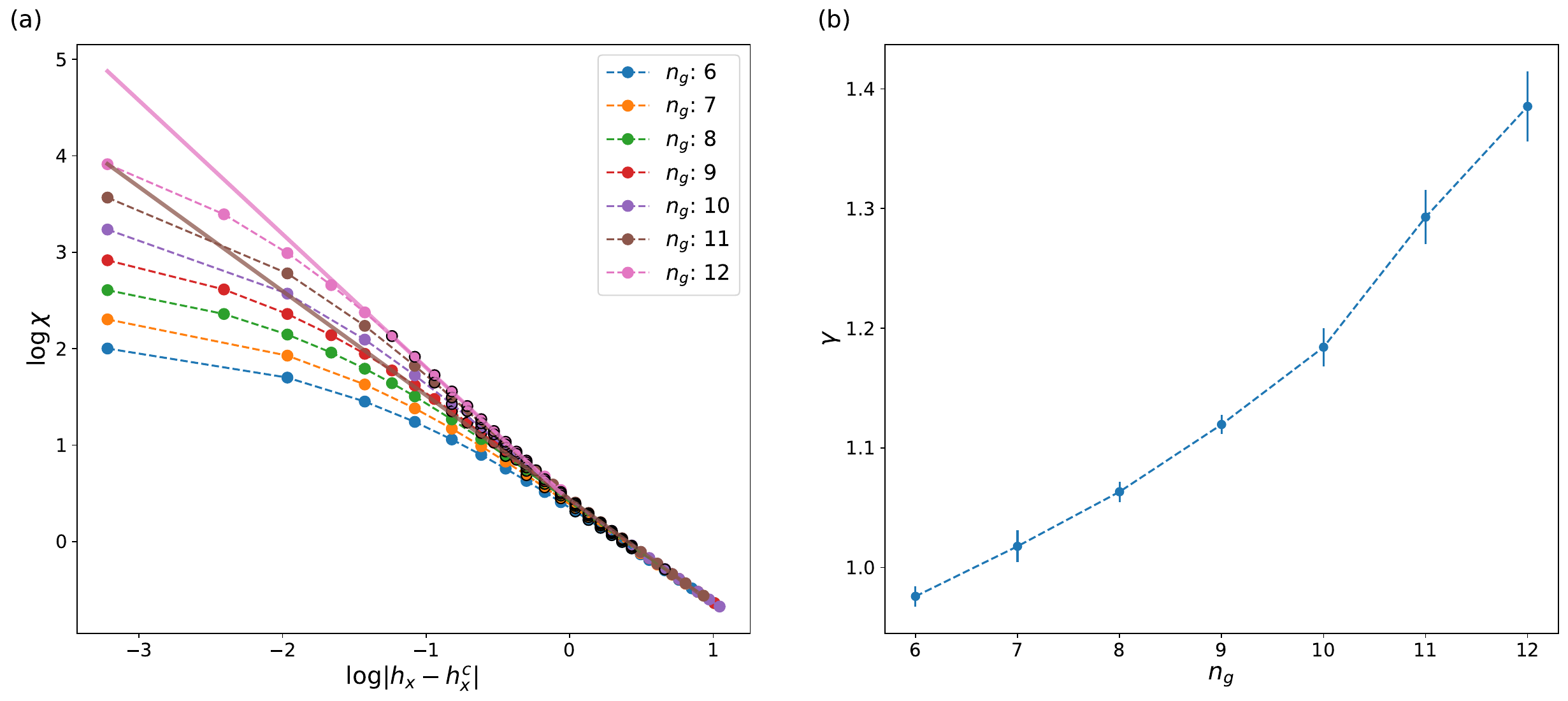}
    \caption{(a) Fit of the magnetic susceptibility in log-log plot for different system size. The pink line highlights the fit used to extract $\gamma$ for $n_g = 12$. The brown line shows the slope of $\chi$ for $n_g = 11$ far from the transition. For each $n_g$ the points highlighted with the black edge are used to extract $\gamma$.  (b) Results of the fit $\gamma$ for different system sizes. The error is calculated as the difference between fit of different data points.}
    \label{fig:fit_gamma_00}
\end{figure}

A more direct observable for characterizing quantum phase transitions on the Cayley tree is the magnetic susceptibility, defined as the response of the magnetization to an applied longitudinal field. Here, we consider the local susceptibility at the central site, which can be expressed as a sum of connected correlations between the central spin and all other sites \cite{Li_efficient}:
\begin{align}
    \chi = \frac{\partial m_z}{\partial h_z} = \sum_{i = 1}^{n_g} N_i\left(\langle \sigma_0^z \sigma_{n_i}^z \rangle - \langle \sigma_0^z\rangle \langle \sigma_{n_i}^z \rangle\right)\,,
    \label{eq:mag_susc}
\end{align}
where $N_i = 3\cdot2^{n_i -1}$ is the number of sites in generation $n_i$.
In the thermodynamic limit, the magnetic susceptibility is expected to diverge algebraically upon approaching the transition\cite{Bose_Hubbard_bethe}:
\begin{align}
    \chi \propto |h_x -h_x^c|^{-\gamma}\, . 
    \label{eq:chi_critical}
\end{align}
At the quantum critical point, the corresponding finite-size scaling with the characteristic scale $N$ is expected to take the form
\begin{align}
    \chi \propto N^{\gamma/\nu}\, . 
    \label{eq:chi_finite}
\end{align}
Fig.\ref{fig:mag_susc_cayley}(a) summarizes our numerical results for the magnetic susceptibility across the transition for different system sizes. On the paramagnetic side of the transition, the magnetic susceptibility shows a clear tendency to diverge upon approaching the finite-size transitions $h_x^f$, with the divergence becoming progressively more pronounced as the size of the tree increases. Indeed, in Fig.\ref{fig:mag_susc_cayley}(b) we uncover a clear algebraic growth of the magnetic susceptibility measured right after the finite size jump with a slope $\theta$:
\begin{equation}
    \chi(h_x^{f+}) \propto N^\theta \propto 2^{\theta n_g}\,.
    \label{eq:chi_max_scaling}
\end{equation}
Comparing the Eqs.\ref{eq:chi_max_scaling} and \ref{eq:chi_finite}, we estimate compute magnetic susceptibility  critical exponent $\gamma=\theta\cdot\nu\approx1.39$, where we used the previously extracted value of $\nu\approx2.63$.
Deeper in the paramagnetic phase, where the dominant correlation scale can be assumed to remain sufficiently small compared with the size of the tree, we can estimate the critical exponent $\gamma$ directly from Eq.\ref{eq:chi_critical}. To this end, we analyze the scaling of the magnetic susceptibility with the distance from the critical point $h_x^c$ on a log-log scale for different tree sizes, as presented in Fig.\ref{fig:fit_gamma_00}(a). The data reveal a crossover, manifested as a change in the slope around $\log|h_x-h_x^c|\approx 0$, which we interpret as the onset of the critical scaling regime. Upon approaching the critical point further, however, finite-size effects become increasingly pronounced. Consequently, the critical scaling can be extracted only within a relatively narrow intermediate window, which gradually broadens with increasing system size.
Tracking the effective slope within this scaling window as a function of system size, we find that the extracted value of $\gamma$ increases monotonously with $n_g$ and does not reach a clear saturation within the accessible system sizes. Our results therefore provide only a lower bound for the critical exponent, $\gamma_\mathrm{min} \sim 1.38$, as shown in Fig.\ref{fig:fit_gamma_00}(b). Nevertheless, for $n_g=12$ we already observe a reasonably broad parameter window beyond the crossover in which the scaling is approximately linear, suggesting that further increases in the tree size may not substantially modify this estimate. Remarkably, this value is also in excellent agreement with the independent estimate of $\gamma$ obtained from the finite-size scaling of the susceptibility at the jump. Unfortunately, the exponential growth of the total number of sites with $n_g$ prevents us from simulating trees with $n_g>12$ generations and from testing this hypothesis directly.

\subsection{Universality}

\begin{figure}[h!]
    \centering
    \includegraphics[width=1\linewidth]{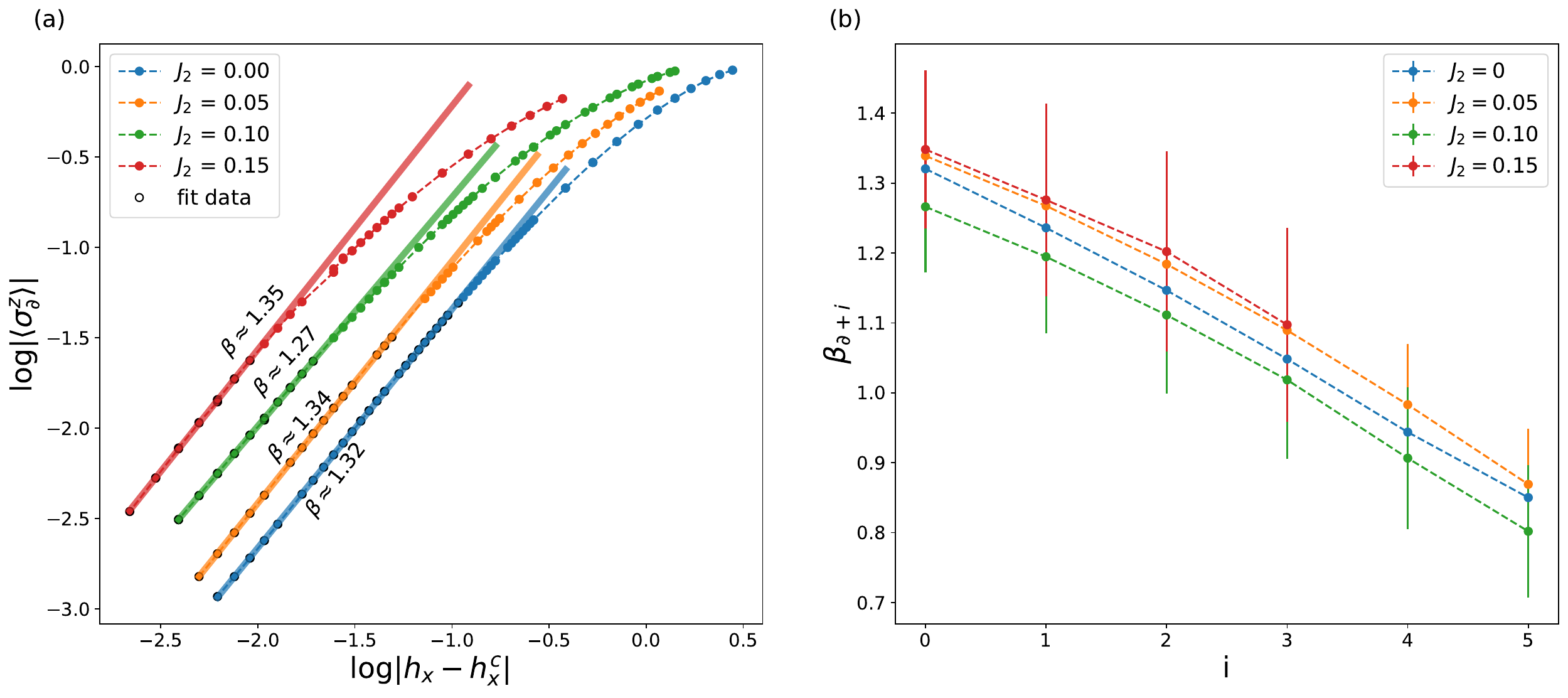}
    \caption{Comparison of the boundary magnetization and $\beta_{\partial-i}$ close to the phase transition. (a) Shows the slope of the boundary magnetization in a log-log plot for different values of the perturbation $J_2$. For each curve the data used for the fit are marked by $\times$. (b) Shows the different values of $\beta_{\partial-i}$ extracted for the different distances and values of $J_2$. The error bars are calculated by shifting the transition point of $\pm 0.1$}
    \label{fig:mag_universality}
\end{figure}

\begin{figure}[h!]
    \centering
    \includegraphics[width=1\linewidth]{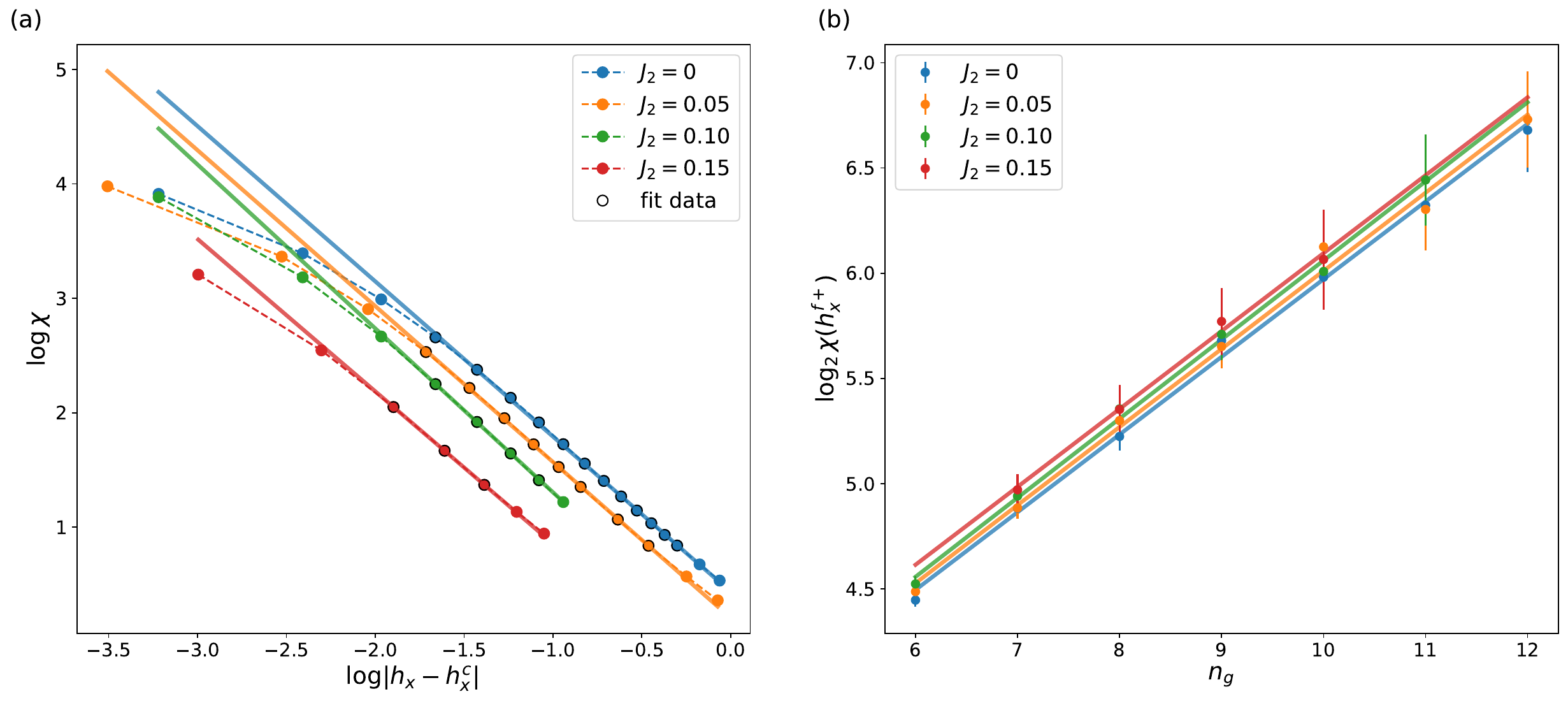}
    \caption{(a) Algebraic scaling of the magnetic susceptibility for different values of $J_2$. We present results for $n_g = 12$ for $J_2\le 0.1$, and for $n_g=11$ for $J_2 = 0.15$. All extracted values of $\gamma$ fall within $5\%$ of $\gamma \approx 1.38$. The points used in the fit are marked with black circles. (b) Algebraic scaling of the maximum magnetic susceptibility $\chi(h_x^{f+})$ for different values of $J_2$. All values extracted fall within $2\%$ of the estimate $\theta \approx 0.53$. }
    \label{fig:univ_gamma_alpha}
\end{figure}

One of the most remarkable properties of continuous phase transitions, quantum or classical, is that their scaling properties are universal. In one and two dimensions, the universality class conjecture allows one to categorize transitions according to the symmetries spontaneously broken on either side of the transition, while the set of critical exponents remains insensitive to the microscopic details of the model.
To the best of our knowledge, the notion of universality classes has not been systematically established for tree lattices or, more generally, for systems defined on curved geometries. Here, we make a first step in this direction. In the previous section, we extracted the critical exponents $\beta$, $\nu$, and $\gamma$ for the transverse-field Ising model. We now test whether the corresponding critical scaling remains unchanged upon deforming the microscopic Hamiltonian by an additional interaction that preserves the $\mathbb{Z}_2$ symmetry. To this end, we consider a tree version of the anisotropic next-nearest-neighbor (NNN) Ising model \cite{ANNNI_1d} introduced previously in Eq.\ref{eq:ANNNI}.
\begin{figure}[h!]
    \centering
    \includegraphics[width=1\linewidth]{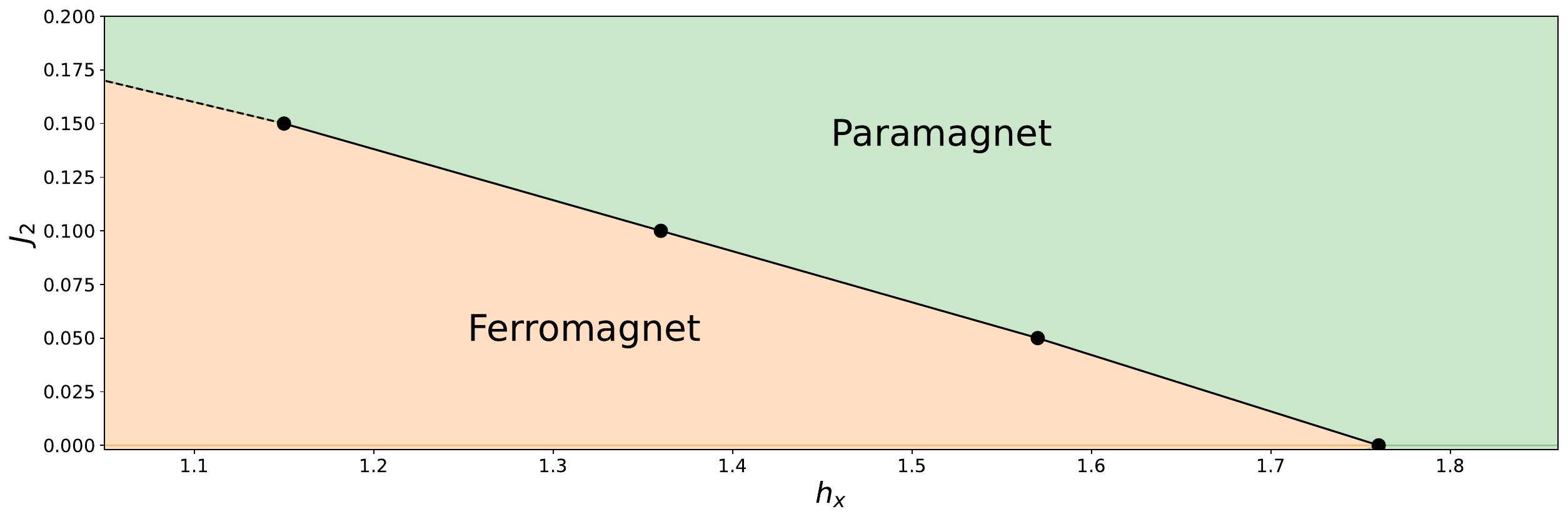}
    \caption{Phase diagram of the anisotropic next-nearest-neighbor Ising model on a tree defined in Eq.\ref{eq:ANNNI}. While the location of the transition shifts with increasing next-nearest-neighbor interaction, the nature and properties of the boundary transition remains unchanged. }
    \label{fig:phase_diagram}
\end{figure}
The next-nearest-neighbor interactions considered here, $J_2 \le 0.15$, do not change the symmetry of the two phases but shift the location of the boundary transition on the Cayley tree, as shown in the phase diagram in Fig.\ref{fig:phase_diagram}. The critical scaling of the order parameter, the boundary magnetization, is shown in Fig.\ref{fig:mag_universality}(a). The extracted values of the critical exponent $\beta$ for different values of $J_2$ agree within $4\%$, providing evidence that the boundary transition remains in the same universality class under this perturbation. The observed variation can be largely attributed to the substantial narrowing of the parameter window over which quantum critical scaling can be reliably extracted for the accessible tree sizes.
Furthermore, we compare how the magnetization critical exponent evolves with the distance from the boundary of a Cayley tree, as shown in Fig.\ref{fig:mag_universality}(b), where we present the values of $\beta_{\partial-i}$ for three different values of the perturbation $J_2$. The first three points at $i=0$ correspond to the three values of $\beta$ extracted in Fig.\ref{fig:mag_universality}(a) for the boundary magnetization. Although the error bars associated with the fitting procedure remain substantial, all three curves exhibit similar behavior. We therefore conclude that, qualitatively, the properties of the boundary transition observed on the Cayley tree remain unaffected by the NNN interaction.
\begin{figure}[h!]
    \centering
    \includegraphics[width=1\linewidth]{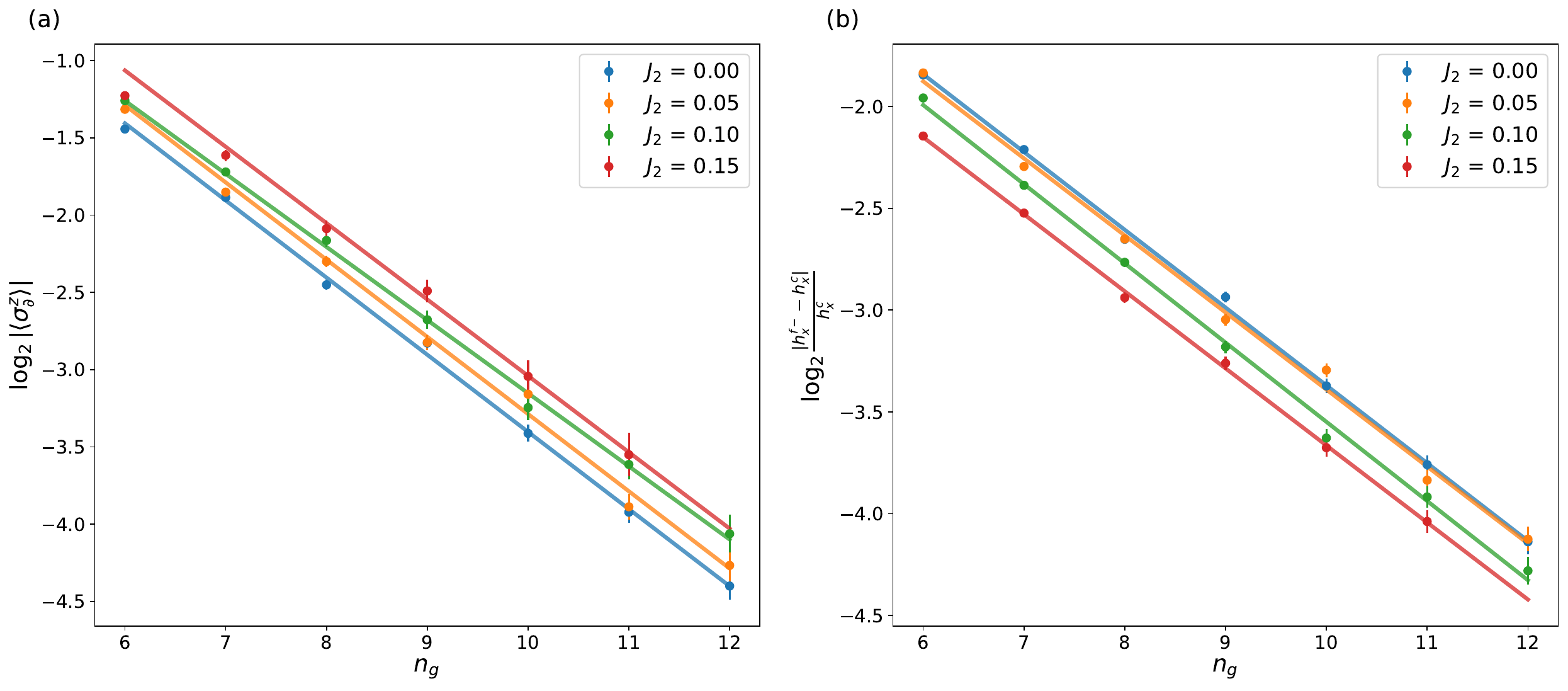}
    \caption{Universal scaling of the (a) magnitude and (b) location of the finite-size jump in the boundary magnetization for different values of the NNN coupling $J_2$. The extracted critical exponents agree with (a) $d\approx0.49$ within 5 \% and (b) $\lambda\approx0.38$ within 2 \%, demonstrating the remarkable robustness of the critical scaling against the ANNNI perturbation.}
    \label{fig:loc_mag_scaling_univ}
\end{figure}
We now compare the finite-size scaling of the location of the finite-size jump and of the boundary magnetization immediately before the jump. In both cases, the scaling at finite $J_2$ closely follows that of the original transverse-field Ising model without NNN perturbations as shown in Fig.\ref{fig:loc_mag_scaling_univ}. The small discrepancies in the numerically extracted values of the scaling dimension $d$ and the critical exponent $\lambda=1/\nu$ are of the order of maximum $5\%$ and can be largely attributed to numerical uncertainties in the fitting. These results support the universality of the previously extracted critical exponents $\beta$ and $\nu$, at least with respect to next-nearest-neighbor perturbations.
Finally, we test the universality of the critical exponent $\gamma$. Following the procedure established above, we extract $\gamma$ directly from the scaling of the magnetic susceptibility upon approaching the transition from the paramagnetic phase, as shown in Fig.\ref{fig:univ_gamma_alpha}(a), and independently extract the ratio $\gamma/\nu$ from the finite-size scaling of the magnetic susceptibility immediately after the jump, as shown in Fig.\ref{fig:univ_gamma_alpha}(b). As noted previously, increasing $J_2$ substantially enhances finite-size effects and progressively narrows the parameter window over which a clear critical scaling can be resolved for the accessible tree sizes. Despite these increasingly strong finite-size effects, the extracted slopes remain remarkably stable, agreeing within $5\%$ over the entire range of $J_2$ considered. 
\label{sec:mag_susc_univ}
The agreement between these two independent scaling analyses provides compelling numerical evidence that the susceptibility exponent $\gamma$, together with the previously established exponents $\beta$ and $\nu$, is universal with respect to NNN perturbations. Our results therefore strongly support the existence of a distinct {\it Ising} universality class associated with the boundary quantum phase transition on the Cayley tree.

\section{Effective Bethe lattice}
\label{sec:EBL}
\begin{figure}[h!]
    \centering
    \includegraphics[width=1\linewidth]{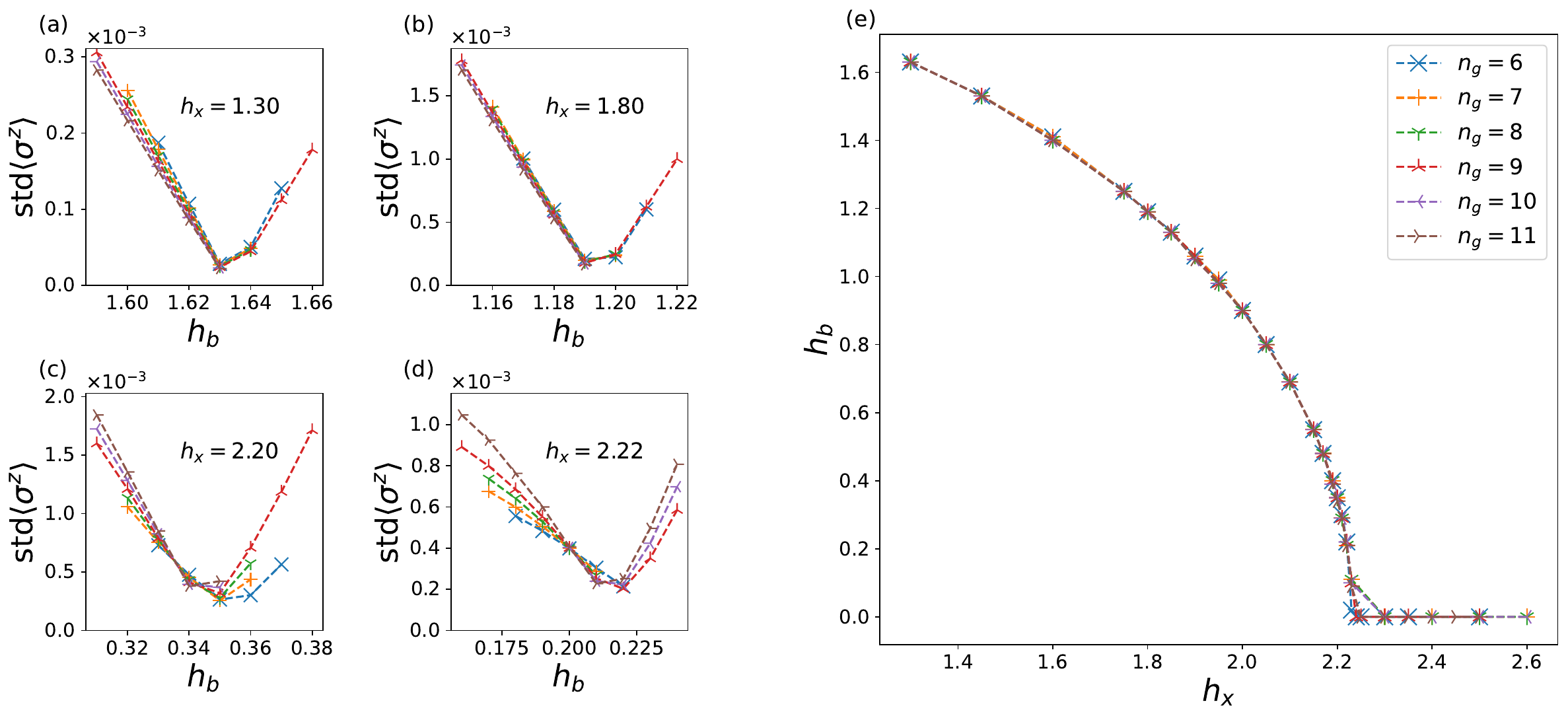}
    \caption{Probing the effective Bethe line. (a)-(d) Minimization of the standard deviation $\mathrm{std} \langle \sigma^z \rangle$ for $n_g = 9$. The plot shows the respectively the results for $h_x = 1.3,\,1.8,\,2.2,\,2.22$. (e) Plot of the effective Bethe line for different system sizes.  Dashed lines are guide to eyes.}
    \label{fig:find_ebl}
\end{figure}

\begin{figure}[h!]
    \centering
    \includegraphics[width=1\linewidth]{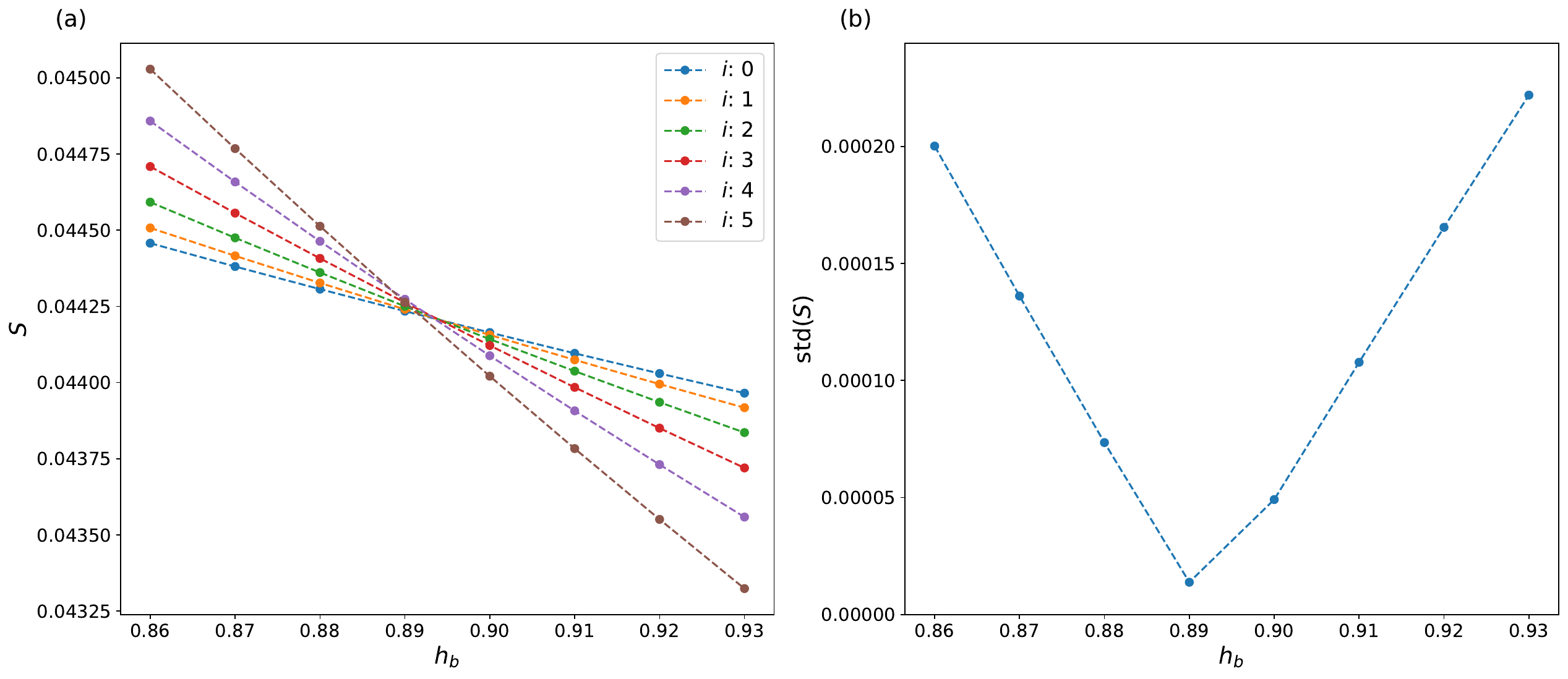}
    \caption{(a) Entanglement entropy across five different bipartition of the Cayley tree at different distances from the center as a function of the boundary field $h_b$ for $h_x = 2$ on a tree of $n_g = 9$. For the first 5 partition starting from the center. (b) Standard deviation of the five sets of  entanglement  data points presented in (a). Both panels suggest that at the effective Bethe line $h_b(h_x=2)\approx0.892$ identified with magnetization as shown in Fig.\ref{fig:find_ebl}, the entanglement is uniform across the Cayley tree.}
    \label{fig:ee_across_ebl}
\end{figure}

\begin{figure}[!ht]
    \centering
    \includegraphics[width=1\linewidth]{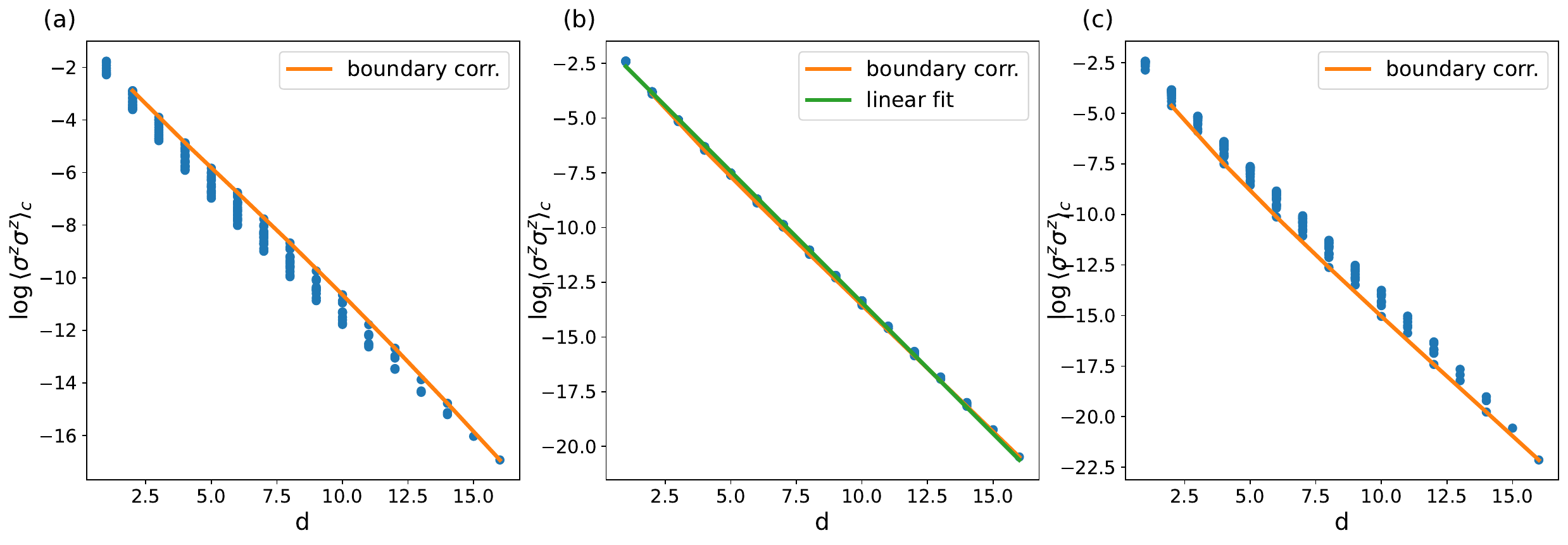}
    \caption{Collapse of the connected correlation functions across the EBL at $h_x = 1.9$. Connected correlations are shown for (a) under-polarized boundary sites, (b) the boundary field $h_b$ fine-tuned to the EBL, and (c) over-polarized boundary sites. In all three panels, we highlight the correlations between boundary sites ($\partial$ corr.) to emphasize the differences between the three regimes. At the EBL, all correlations at a given graph distance collapse onto a single curve.  }
    \label{fig:cc_crossing}
\end{figure}

\begin{figure}[!ht]
    \centering
    \includegraphics[width=1\linewidth]{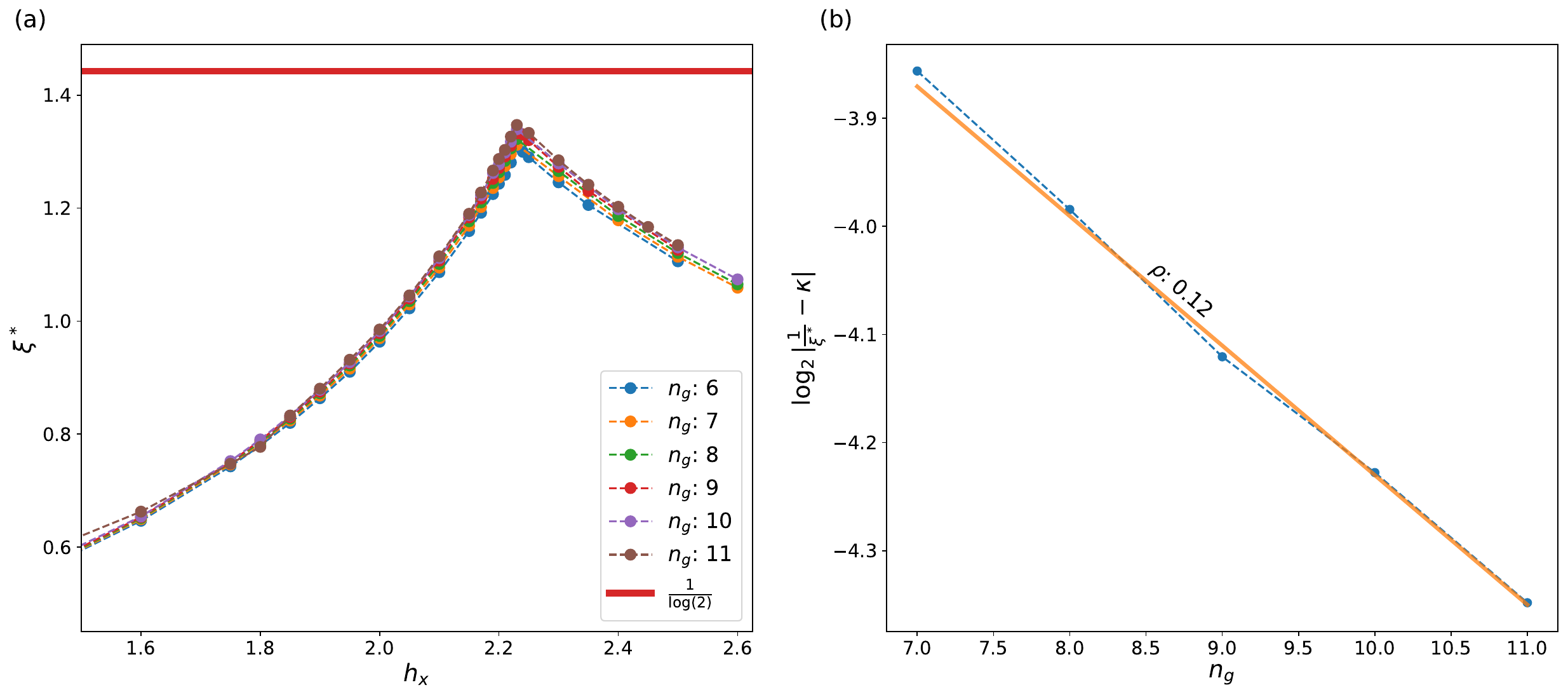}
    \caption{ (a) Effective correlation length $\xi^*$ as a function of the transverse field for finite-size Cayley trees with $n_g$ generations, fine-tuned to the effective Bethe line. The effective correlation length is bounded by the curvature scale of the lattice, $1/\log(2)$ (red line), while for finite trees it is further suppressed by finite-size effects. (b) Finite-size scaling of the measured inverse correlation length $1/\xi^*$ after subtracting the curvature contribution, evaluated at the critical point $h_x \approx 2.23$. The scaling exponent $\rho\approx0.12$ quantifies the leading finite-size correction to the correlation length extracted from a fine-tuned Cayley tree.} 
    \label{fig:cc_ebl}
\end{figure}

The fundamental difference between the infinite Bethe lattice and a finite Cayley tree lies in the presence of a boundary. Locally, the two lattices have the same coordination and branching structure, but the Bethe lattice is infinite and boundary-free, whereas the Cayley tree contains an extensive number of boundary sites. As demonstrated in the previous sections, this distinction has profound consequences for quantum criticality: while the transverse-field Ising model on the Bethe lattice undergoes a mean-field transition, the finite Cayley tree exhibits a distinct boundary-driven transition at a different critical point and with a different set of critical exponents.
This observation naturally raises the question of whether the Bethe-lattice critical behavior can be recovered on a finite Cayley tree by suppressing the influence of its boundary. Importantly, the mean-field nature of the transition on the infinite Bethe lattice suggests that such a construction may be possible: mean-field behavior is determined by an effective local field generated by the surrounding environment, implying that the effect of the outer branches extending beyond the boundary of the Cayley tree may, in principle, be reproduced by an appropriate local field applied to the boundary sites. If such a compensation is possible, we should be able to suppress the boundary transition discussed above and recover the bulk mean-field transition of the infinite Bethe lattice.
We therefore introduce a longitudinal boundary field $h_b$:
\begin{equation}
    \mathcal{H}_{Bethe} = \mathcal{H}_I - h_b \sum_{i \in \partial} \sigma_i^z\,,
    \label{eq:EBL}
\end{equation}
where $\mathcal{H}_I$ is the transverse-field Ising model (see Eq.\ref{eq:TFIM} for definition). By tuning the strength $h_b$ of the boundary field, we identify a line in the parameter space along which the boundary effects are screened. We refer to this line as the \textit{effective Bethe line}. Along this line, despite retaining the finite Cayley-tree geometry with an extensive physical boundary, the system recovers the critical properties of the boundary-free infinite Bethe lattice.
In order to identify the boundary field that effectively screens the presence of the boundary, we examine the variation of the local magnetization across different generations. On the infinite Bethe lattice, the magnetization is spatially uniform, and therefore its standard deviation across the tree should vanish. We thus identify the effective Bethe line by tuning $h_b$ to minimize the variation of the magnetization between different generations. Examples for the tree with $n_g=9$ generations and various transverse fields are shown in Fig.\ref{fig:find_ebl} (a)-(d). Owing to the finite numerical resolution, we accept points with $\mathrm{std}\langle \sigma_i^z \rangle < 5\cdot 10^{-3}$ as being sufficiently close to the effective Bethe line for practical purposes, while always taking the point with the smallest standard deviation as our best estimate. We find that the boundary field required to screen the boundary decreases monotonically with increasing transverse field and exhibits essentially no finite-size dependence, except in the vicinity of the critical point, there finite-size effects become more pronounced and are analyzed in detail in App.\ref{app:EBL_trans}. The resulting effective Bethe line is summarized in Fig.\ref{fig:find_ebl}(e).
An additional check comes from the analysis of the entanglement entropy across the tree, which can be directly extracted from the singular values of the TTNS. Cutting any bond of the tree defines a bipartition into two disconnected branches, and the corresponding von Neumann entanglement entropy is given by
\begin{equation}
    S=-\sum_{\alpha}\lambda_{\alpha}^2\log\lambda_{\alpha}^2\,,
\end{equation}
where $\lambda_{\alpha}$ are the singular values associated with the corresponding bond. On the infinite Bethe lattice, translational invariance implies that the entanglement entropy is uniform across equivalent bonds of the tree. Therefore, along the effective Bethe line on a finite Cayley tree, we expect the entanglement entropy associated with different branch bipartitions to become uniform, providing an independent measure of the screening of boundary effects. In Fig.\ref{fig:ee_across_ebl}(a), we show the entanglement entropy for a finite Cayley tree with $n_g = 9$ generations as a function of the boundary field, evaluated across five cuts located at different distances from the center of the tree. Remarkably, all five entanglement entropies cross at the boundary field independently identified from the uniformity of the magnetization as the effective Bethe line. At this crossing, the entanglement entropy becomes independent of the position of the cut, demonstrating that the boundary field restores the spatially uniform entanglement structure characteristic of the infinite Bethe lattice. The standard deviation measured over these five five bi partitions is presented in Fig.\ref{fig:ee_across_ebl}(b) and goes to zero at the effective Bethe line.  
Finally, we examine whether a Cayley tree with a fine-tuned boundary field can reproduce the quantum phase transition of the infinite Bethe lattice. We first consider the scaling of the order parameter. Since the effective Bethe line tunes the Cayley tree to a regime in which the magnetization is spatially uniform, the order parameter can be identified with the local magnetization at any site of the tree. For concreteness, we choose the magnetization at the central site, $\langle\sigma_0^z\rangle$.
\begin{figure}[!ht]
    \centering
    \includegraphics[width=1\linewidth]{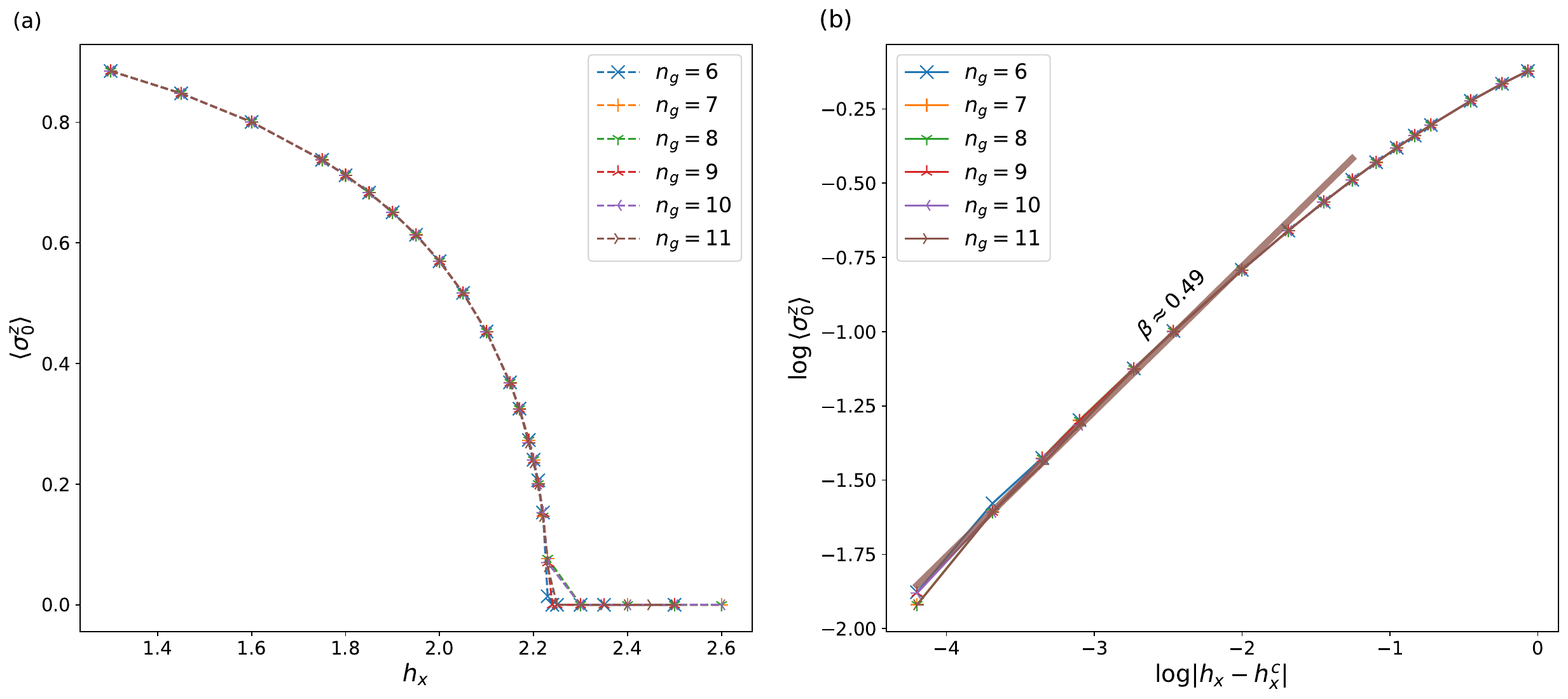}
    \caption{ (a) Local magnetization as a function of the transverse field along the effective Bethe line. (b) Scaling of the local magnetization shown in (a) with the distance from the critical point $h^c_x\approx 2.235$. The numerically extracted critical exponent is in excellent agreement with the mean-field value $\beta=1/2$. }
    \label{fig:mag_EBL}
\end{figure}
From the  scaling of the order parameter we locate the critical value along the EBL at $h_x^B \sim 2.235$, which is compatible within 0.2 $\%$ with previous results \cite{TFIM_Nagaj,Li_efficient}.  Numerically extracted order parameter critical exponent $\beta\approx0.49$ agrees within $2\%$ with the mean-field value $1/2$.
Interestingly enough, the existence of the effective Bethe line is closely tied to the mean-field nature of the transition, while remaining largely insensitive to the specific geometry of the tree. For instance, in App.\ref{app:ebl_4} we demonstrate the existence of an effective Bethe line and recover the mean-field scaling of the order parameter for the transverse-field Ising model on a tree with a different branching number, $q=4$.
In striking contrast to the Cayley tree, where the strong boundary effects prevent us from extracting a well-defined correlation length from the connected correlations (Fig.\ref{fig:CC_cayley}), along the EBL the correlation length can be extracted even for relatively small systems. As a consequence of the fine-tuned boundary condition, the connected correlations ``collapse" onto a single curve and exhibit a clear exponential decay with the graph distance $d$.
In Fig.\ref{fig:cc_crossing}, we show representative correlation functions for different values of the boundary field: fine-tuned to the effective Bethe line, under-polarized, and over-polarized. To highlight the difference between the under- and over-polarized systems and the EBL, we again emphasize the correlations along the boundary ($\partial$ corr.), as in Fig.\ref{fig:CC_cayley}. At the fine-tuned boundary field, we observe a remarkable collapse of all connected correlations at a given graph distance $d$, demonstrating that the dependence on the position of the two sites relative to the boundary is effectively eliminated. The correlations are therefore determined solely by their graph distance, as expected for the homogeneous infinite Bethe lattice.
Remarkably, the connected correlations decay exponentially with $d$ even in the vicinity of the critical point. This behavior is in striking contrast to the algebraic decay characteristic of criticality in one dimension, but is fully consistent with previous studies of the infinite Bethe lattice \cite{Li_efficient}. Additional numerical data for the connected correlations along the effective Bethe lines can be found in App.\ref{app:cc_EBL}.
The correlation length extracted along the effective Bethe line on finite Cayley trees is summarized in Fig.\ref{fig:cc_ebl}(a). Fine-tuning the boundary field to the effective Bethe line does not completely eliminate finite-size effects from non-local observables, which remain visible through subleading finite-size corrections. Nevertheless, despite the slow growth of the extracted correlation length with increasing tree size, it is evident that the observed behavior is far from a conventional divergence. In the literature \cite{Li_efficient,Samerjian_bose}, this peculiar feature has been discussed in connection with the divergence of the magnetic susceptibility. Here, we propose an alternative geometric interpretation.
Following Mosseri and Sadoc\cite{mosseri}, we view the Bethe lattice as a limiting tessellation of hyperbolic space with Schl\"{a}fli symbol $\{\infty,q\}$, where $q$ is the coordination number of the tree. In contrast to flat space, hyperbolic space possesses an intrinsic length scale set by its curvature $\kappa$, which profoundly modifies the spatial decay of correlations. In particular, it has recently been shown that correlations in hyperbolic space differ qualitatively from those in flat space \cite{critical_hyperbolic}:
\begin{equation}
    \langle \sigma_{p_i}^z \sigma_{p_j}^z \rangle_c \propto \frac{e^{-\frac{d}{\xi}}}{\sinh(\kappa d)}\,,
    \label{eq:hyper_corr}
\end{equation}
where $\xi$ is the intrinsic correlation length, which diverges at criticality, while $\kappa$ remains finite for a given lattice geometry. For the Bethe lattice, $\kappa = 1/\log(q-1)$, and therefore for $q=3$, $\kappa = 1/\log2$. Consequently, the effective correlation length extracted from the exponential decay of the numerical data does not measure $\xi$ alone, but combines the critical correlation scale with the intrinsic geometric scale set by the curvature.
Therefore, when the system approaches the critical point correlation length and curvature start to compete up to the critical point where only the curvature contribution remains:
\begin{equation}
    \langle \sigma_{p_i}^z \sigma_{p_j}^z \rangle_c \propto e^{-\left(\frac{1}{\xi} + k \right) d } \propto e^{-\frac{d}{\xi^*}} \,.
    \label{eq:cc}
\end{equation}
For this reason the extraction of the critical exponent $\nu$ is quite challenging, as in the region near the transition where one would normally extract this critical  exponent there is a strong competition between the two different length scales.
We find that, in addition to the curvature contribution, the correlation length extracted from finite Cayley trees exhibits a systematic suppression due to finite-size effects:
\begin{equation}
    \left|\frac{1}{\xi^*} - \kappa\right| = \frac{1}{\xi} \propto N^{-\rho}\,.
\end{equation}
The finite-size scaling of the curvature-corrected inverse correlation length at the critical point is shown in Fig.\ref{fig:cc_ebl}(b), revealing a clear algebraic behavior governed by the exponent $\rho\approx0.12$. Whether this finite-size correction, and in particular the exponent $\rho$, is universal across different realizations of effective Bethe lattices remains an interesting question for future studies.

\section{Conclusion and discussion}
\label{sec:conclusions}

In this work, we have investigated the quantum phase transition of the transverse-field Ising model on a finite Cayley tree. Our results demonstrate that the presence of an extensive boundary fundamentally changes the critical behavior compared with the infinite, boundary-free Bethe lattice. While the latter exhibits the conventional mean-field transition, on the Cayley tree we find a distinct boundary-driven quantum phase transition, occurring at $h_x^c\approx1.76$, well below the Bethe-lattice critical point $h_x^c\approx2.23$. This difference persists in the thermodynamic limit because the fraction of boundary sites remains finite as $n_g\rightarrow\infty$. Consequently, increasing the number of generations does not continuously transform a Cayley tree into the infinite Bethe lattice - the existence of a boundary remains a fundamental distinction between the two lattices, even in the thermodynamic limit.

\noindent
For any finite tree, the transition manifests itself as a discontinuous jump of the magnetization. Both the magnitude and the location of this jump, however, exhibit systematic finite-size scaling: the jump vanishes while its position converges towards $h_x^c\approx1.76$ with increasing $n_g$. Our results indicate that the first-order transition at finite sizes evolves into a continuous boundary quantum phase transition in the thermodynamic limit. The boundary magnetization naturally emerges as the corresponding order parameter. Its critical scaling yields a boundary exponent $\beta_\partial\approx 1.32$, while the effective exponent extracted at successive generations decreases upon moving away from the boundary, reflecting a gradual crossover from boundary-dominated to bulk (mean-field like) behavior.

\noindent
The unusual geometry of the Cayley tree requires a finite-size scaling framework different from that of conventional finite-dimensional lattices. Since $N$ grows exponentially with the number of generations, scaling that is exponential in $n_g$ becomes algebraic in $N$. From the finite-size drift of the transition, we identify a characteristic exponent $\lambda\approx0.38$, corresponding within our interpretation to $\nu=1/\lambda\approx2.63$. The independently extracted scaling dimensions of the local magnetization satisfy $d_i=\beta_i/\nu$ within a few percent. The magnetic susceptibility provides an additional and independent characterization of the transition. Its finite-size scaling yields $\gamma/\nu\approx0.53$ and, consequently, $\gamma\approx1.39$. A direct analysis of the susceptibility upon approaching the transition from the paramagnetic phase gives a consistent estimate. Together, these results establish a coherent set of critical exponents characterizing the boundary Ising transition on the Cayley tree.

\noindent
A particularly important question is whether these exponents describe universal critical behavior rather than a specific microscopic realization of the transverse-field Ising model. We addressed this question by introducing anisotropic next-nearest-neighbor interactions that preserve the $\mathbb{Z}_2$ symmetry. Despite increasingly strong finite-size effects, the critical scaling of the boundary magnetization, the finite-size drift of the transition, and the magnetic susceptibility remain remarkably stable over the range of perturbations considered. The independently extracted values of $\beta$, $\nu$, and $\gamma$ remain consistent with those of the unperturbed model. Our results therefore provide strong numerical evidence for a distinct Ising universality class on the Cayley tree.

\begin{figure}[h!]
    \centering
    \includegraphics[width=0.5\linewidth]{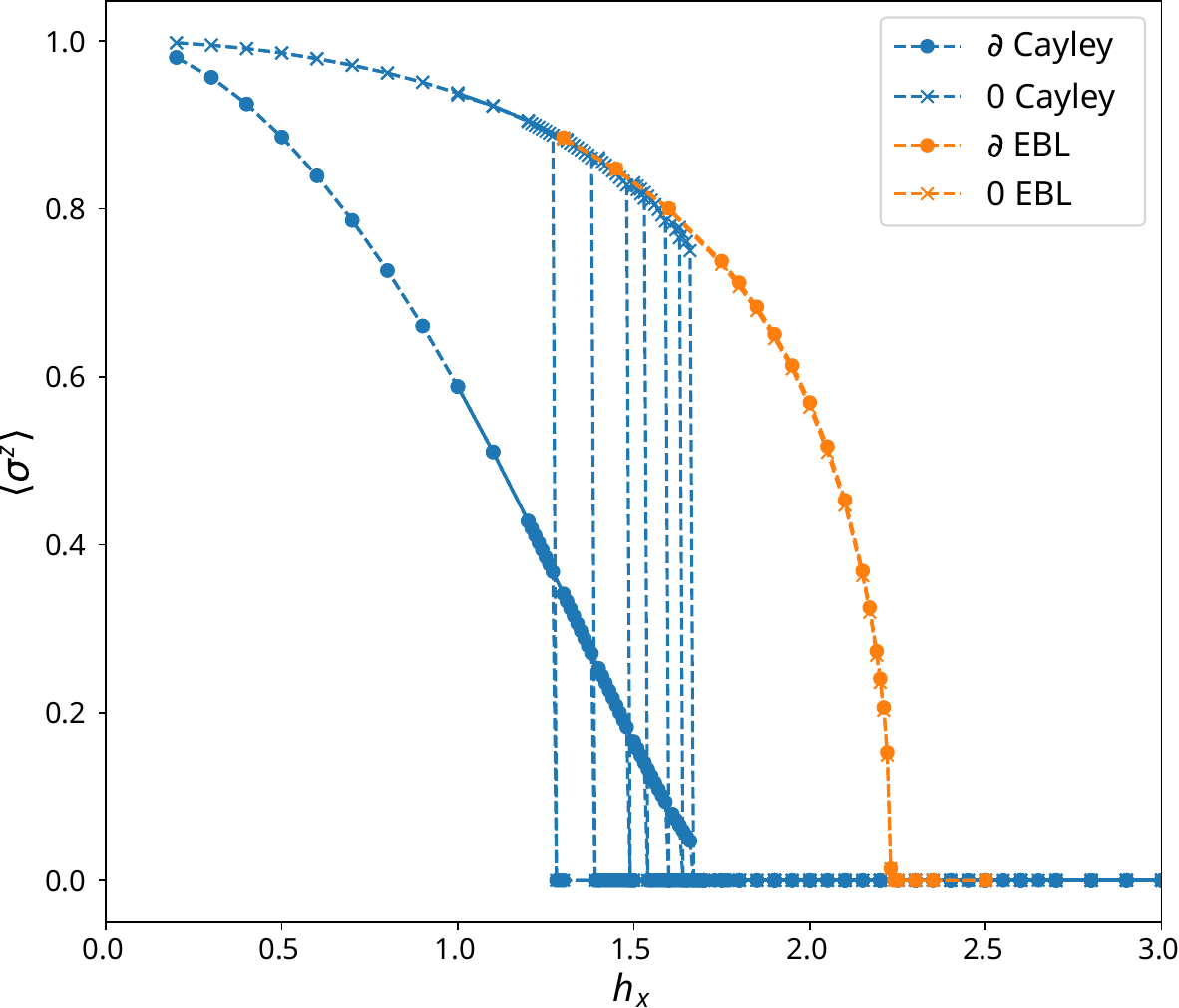}
    \caption{ Local magnetization for the Effective Bethe lattice (orange) and the Cayley tree (blue). Circles and crosses mark magnetization at the boundary and in the center correspondingly.}
    \label{fig:comparison}
\end{figure}

\noindent
Remarkably, the boundary-driven critical behavior can be eliminated by applying an appropriately fine-tuned longitudinal field to the boundary sites. The existence of such an effective boundary field is intimately connected to the mean-field nature of the bulk Bethe-lattice transition: the effect of the environment missing beyond the boundary can be reproduced by a local field acting on the boundary sites. Along the resulting \textit{effective Bethe line}, the magnetization and entanglement entropy become spatially uniform, and the finite Cayley tree recovers the mean-field transition of the infinite Bethe lattice at $h_x^c\approx2.23$ with $\beta=1/2$. The striking difference between the boundary-driven transition of the Cayley tree and the mean-field transition recovered along the effective Bethe line is summarized in Fig.\ref{fig:comparison}. The existence of the effective Bethe line also for a different branching number, $q=4$, further demonstrates that this construction is not specific to a particular tree geometry, but is a robust consequence of the mean-field nature of the corresponding Bethe lattice.

\noindent
In summary, our results reveal two fundamentally distinct manifestations of quantum criticality on tree lattices: a boundary-driven transition intrinsic to finite Cayley trees and the bulk mean-field transition of the infinite Bethe lattice, which can be recovered on the same finite tree by screening its boundary. Thus, unlike in conventional boundary critical phenomena, the boundary of a tree does not merely introduce subleading corrections but can determine both the location and the universality class of the quantum phase transition. This distinction is particularly relevant for experimental realizations of finite tree geometries, where increasing the system size does not eliminate the extensive boundary. At the same time, our results demonstrate that local control of the boundary provides a direct way to tune between boundary-dominated and bulk mean-field quantum criticality within the same finite system.

\section{Acknowledgments}
This work was supported by Kavli innovation award, by the Royal Society (grant number URFR1251326) and  by the European Union through the ERC grant (TRANGINEER, 101220181). The views and opinions expressed are those of the authors only and do not necessarily reflect those of the European Union or the European Research Council Executive Agency; neither the European Union nor the granting authority can be held responsible for them. Numerical simulations were performed at the DelftBlue HPC and at
the Dutch national e-infrastructure with the support of the SURF Cooperative.

\newpage
\clearpage
\begin{appendix}
    \section{Additional data for magnetization on the Cayley tree}
    \subsection{Magnetization distribution across the tree}
    \label{app:mag_across_tree}
   In the main text, we characterize the local magnetization using a single representative site per generation, assuming the symmetry of the Cayley-tree ground state. Here, we provide numerical evidence supporting this assumption.
    \begin{figure}[h!]
        \centering
        \includegraphics[width=1\linewidth]{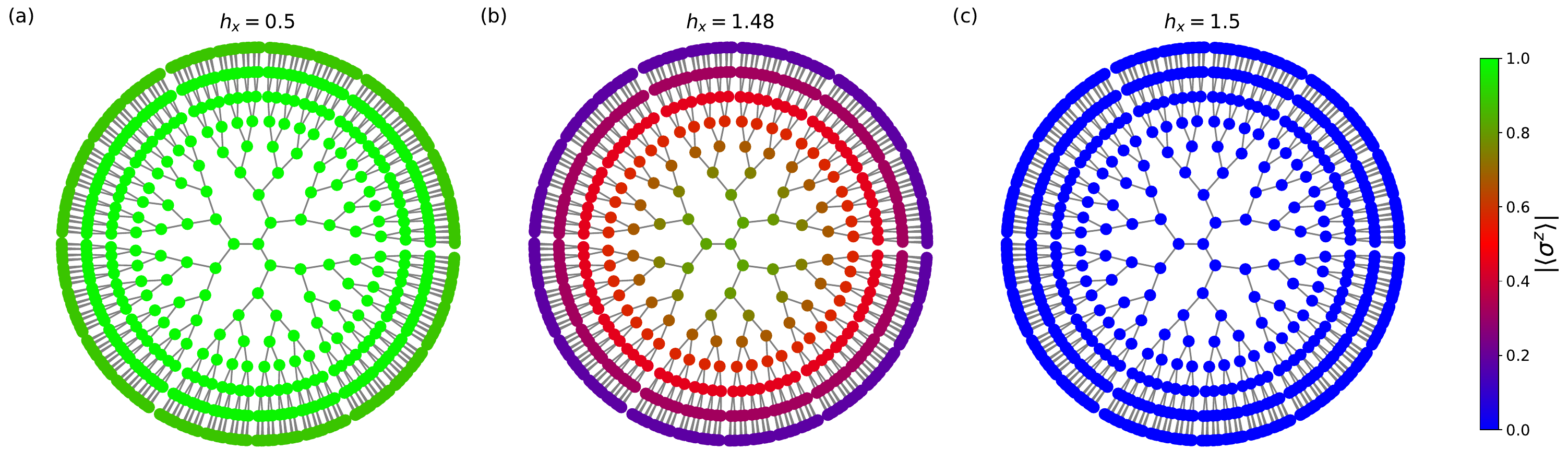}
        \caption{Local magnetization profile over a Cayley tree with $n_g = 8$ (a) deep inside the ferromagnetic phase at $h_x = 0.5$, (b) just before the finite-size jump at $h_x=1.48$; and (c) deep inside the paramagnetic phase at $h_x=1.5$.}
        \label{fig:local_mag_across}
    \end{figure}
\noindent
In addition, the profile shown in Fig.\ref{fig:local_mag_across}(b) clearly demonstrates that the paramagnetic phase first develops at the boundary and progressively penetrates into the bulk. This provides further support for our interpretation of the transition observed on a finite Cayley tree as a {\it boundary} Ising transition.

    \subsection{Boundary and near-boundary magnetization}
    \label{app:dist_bound}

    \begin{figure}[h!]
        \centering
        \includegraphics[width=1\linewidth]{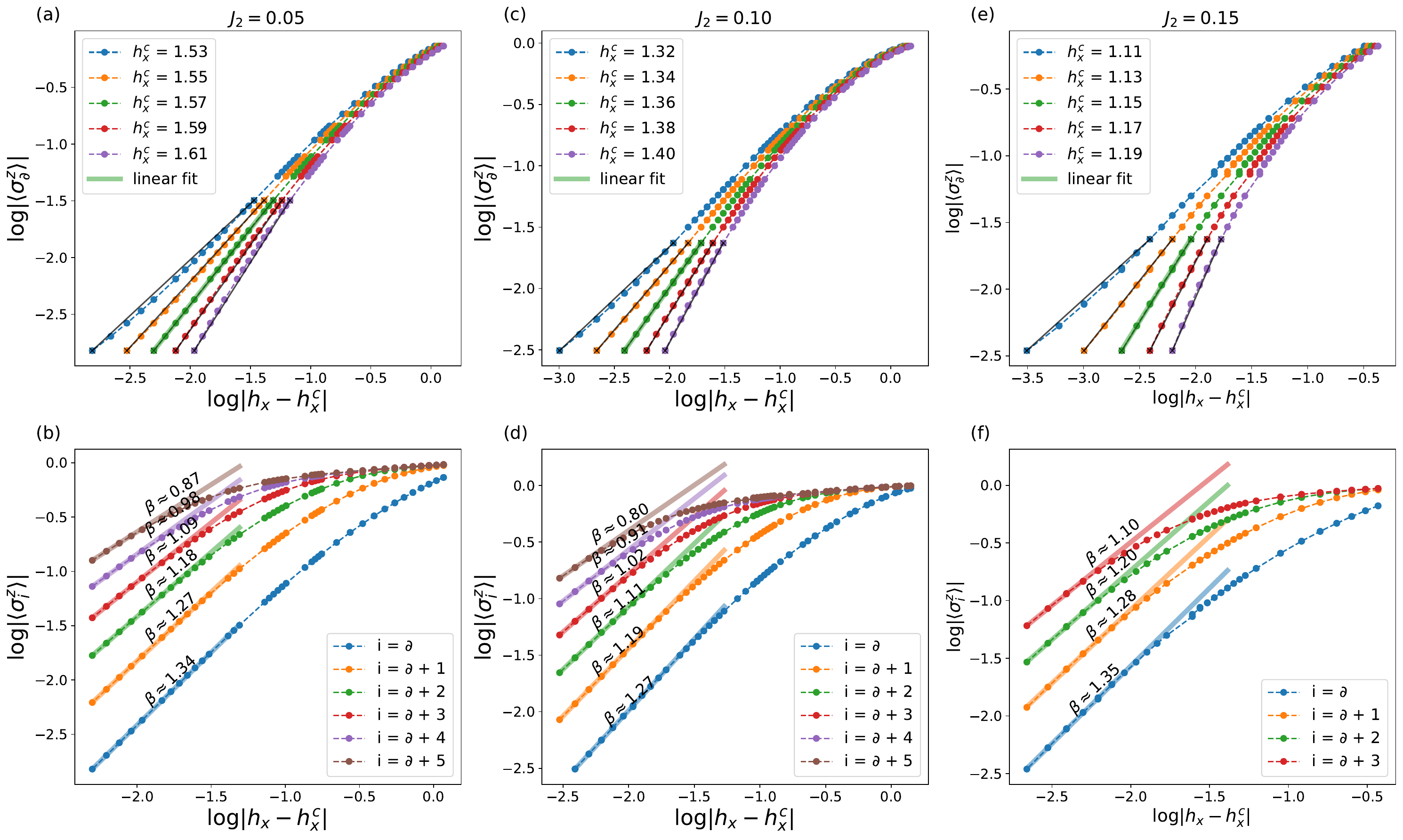}
        \caption{
        Scaling of the boundary magnetization of the ANNNI model on a Cayley tree. Top panels: We identify the critical point with  the separatrix (green) in the log-log plot of the boundary magnetization versus the distance from the transition for (a) $J_2=0.05$, (c) $J_2=0.1$, and (e) $J_2=0.15$. Bottom panels: Scaling of the local magnetization at different distances from the boundary ($\partial + 0$ denotes the boundary sites), with $h_x^c$ fixed to the values determined in the top panels for (b) $J_2=0.05$, (d) $J_2=0.1$, and (f) $J_2=0.15$. The extracted critical exponents are listed near the corresponding data.}
        \label{fig:local_mag_annni}
    \end{figure}
\begin{figure}[h!]
        \centering
        \includegraphics[width=1\linewidth]{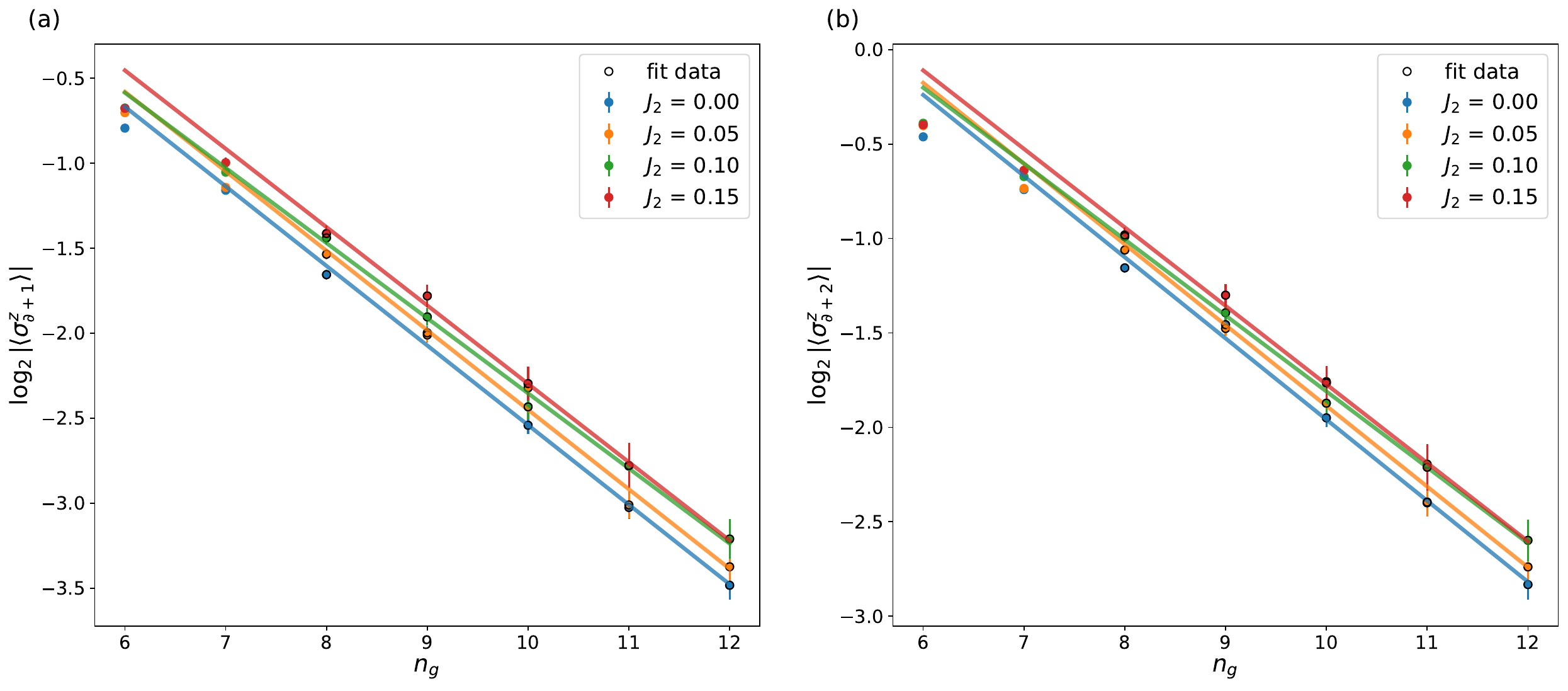}
        \caption{Scaling of the local magnetization for different ANNNI perturbations at distance (a) one and (b) two generations away from the boundary. In both cases the extracted critical exponents agree with each other within 6 \%. Points used in the fitting are marked by black circles.}
        \label{fig:mag_scaling_bulk}
    \end{figure}

    In Fig.\ref{fig:local_mag_annni}(a), (c), (e), we show how the critical points are identified for $J_2 = 0.05$, with $h_x^c \approx 1.57$; and $J_2 = 0.10$, with $h_x^c \approx 1.36$; and  and $J_2 = 0.15$, with $h_x^c \approx 1.15$ correspondingly. The uncertainty of our estimates is approximately $0.01$, which appears sufficient to test the universality of the critical exponents discussed in the main text.
    In Fig.\ref{fig:local_mag_annni}(b), (d), (f), we show the scaling of the local magnetization with the distance to the transition used to numerically extract the exponents $\beta_i$ shown in Fig.\ref{fig:mag_universality}(b) in the main text. The parameter window over which a reliable fit can be performed decreases as the sites approach the center of the tree.
    In the main text we reported we focus our discussion of universality on the boundary magnetization. Here we compare the scaling of local magnetization one and two generations away from the boundary. As with the boundary magnetization, finite-size scaling appears to be universal with respect to the ANNNI perturbations $J_2$ as shown in Fig.\ref{fig:mag_scaling_bulk}.

    \newpage
    \section{Effective Bethe line near the transition}
    \label{app:EBL_trans}

\begin{figure}[h!]
        \centering
        \includegraphics[width=1\linewidth]{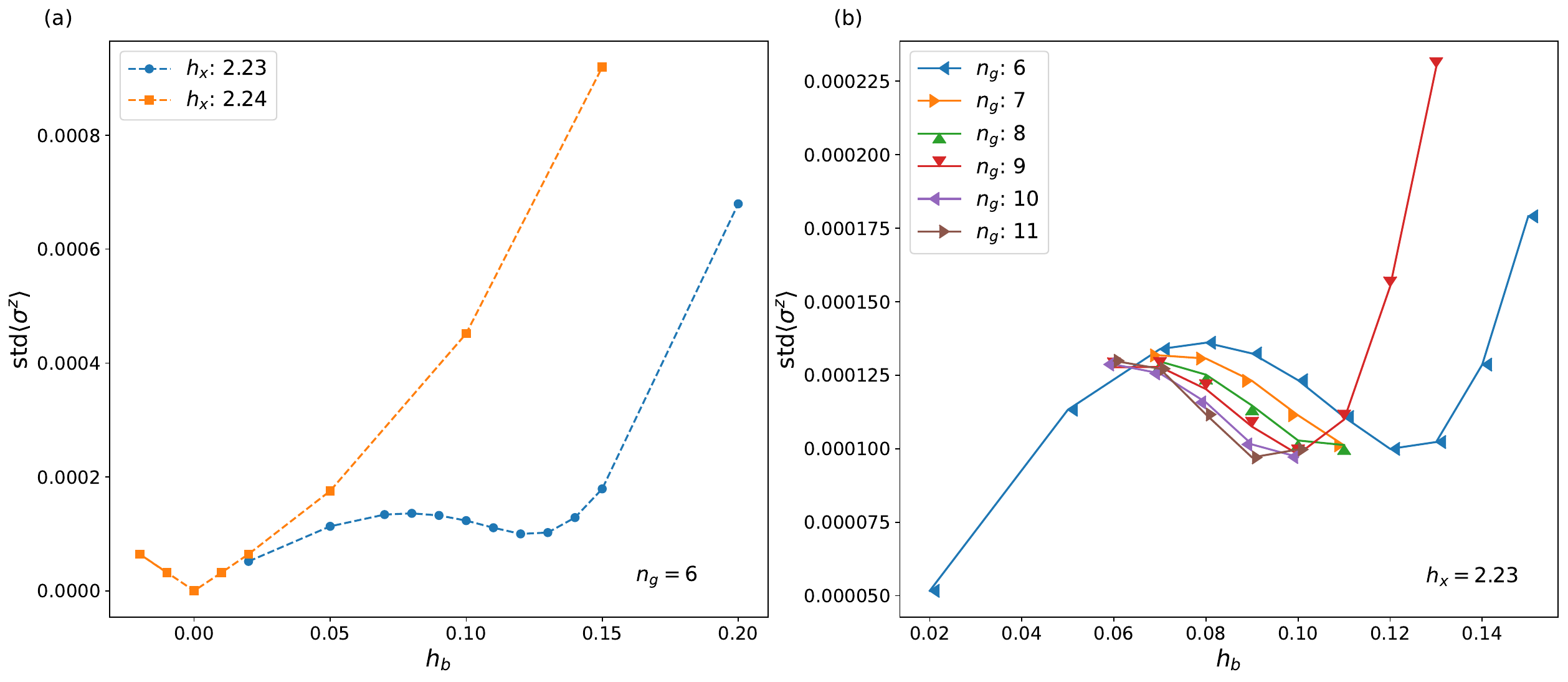}
        \caption{Behavior of the standard deviation close to the transition. (a) Shows the different behavior of $\mathrm{std}\langle \sigma^z\rangle$ on the ferromagnetic (circle) and paramagnetic (square) sides of the transition. (b) Shows the slope of $\mathrm{std}\langle \sigma^z\rangle$ for different system sizes at $h_x = 2.23$, on the ferromagnetic side of the transition. }
        \label{fig:ebl_close}
    \end{figure}

    The Effective Bethe line allows us to find the transition point on the Bethe lattice , which differ substantially from the Cayley tree. In the Cayley tree in the paramagnetic phase the magnetization is uniform within our threshold of $0.0005$ before we reach the transition points of the Bethe lattice. These points could be naively considered belonging to the EBL, but there is a considerable difference in the behavior of the magnetization in the two phases.
    In Fig.\ref{fig:ebl_close}(a) we show the different behavior of $\mathrm{std}\langle\sigma^z\rangle$ on the two sides of the transition on the Bethe lattice. In the paramagnetic side (orange line) we can clearly see that as we decrease the boundary field $h_b$, $\mathrm{std}\langle\sigma^z\rangle$ decreases monotonically. Until it reaches its minimum at $h_b = 0$. Differently on the ferromagnetic side of the transition $\mathrm{std}\langle\sigma^z\rangle$ doesn't it is not monotonous. Around $h_b \approx 0.12$ we find a local minimum. This allows us to safely determine which points fall onto the effective Bethe line. All points that exhibit a local minimum only for $h_b=0$, will be in the paramagnetic phase, while all points that exhibit a local minimum at $h_b\ne0$ belong to the effective Bethe line.
    In the main text we argued that the effective Bethe line shows no significant finite-size dependence. However, closer to the transition the EBL start to drift. In Fig.\ref{fig:ebl_close}(b) we show the minimization of $\mathrm{std}\langle \sigma^z\rangle$ at $h_x =2.23$ for $n_g = 6,7,8,9,10,11$. In the plot we clearly see a  shift of the EBL slowly converging upon increasing the size of the tree, while the local  minimum remains at finite values of $h_b > 0$. Taking this into account we are still able to estimate the EBL close the the transition for different system sizes.

    \section{Effective Bethe line for $q=4$}
    \label{app:ebl_4}
    Here we identify the EBL in trees with coordination number $q =4$.  We follow the same procedure by looking at the vanishing standard deviation of the magnetization across the tree. The transition along the effective Bethe line is consistent with mean-field type with $\beta=1/2$ and takes place at $h_x^c\approx2.29$.
    \begin{figure}[h!]
        \centering
        \includegraphics[width=1\linewidth]{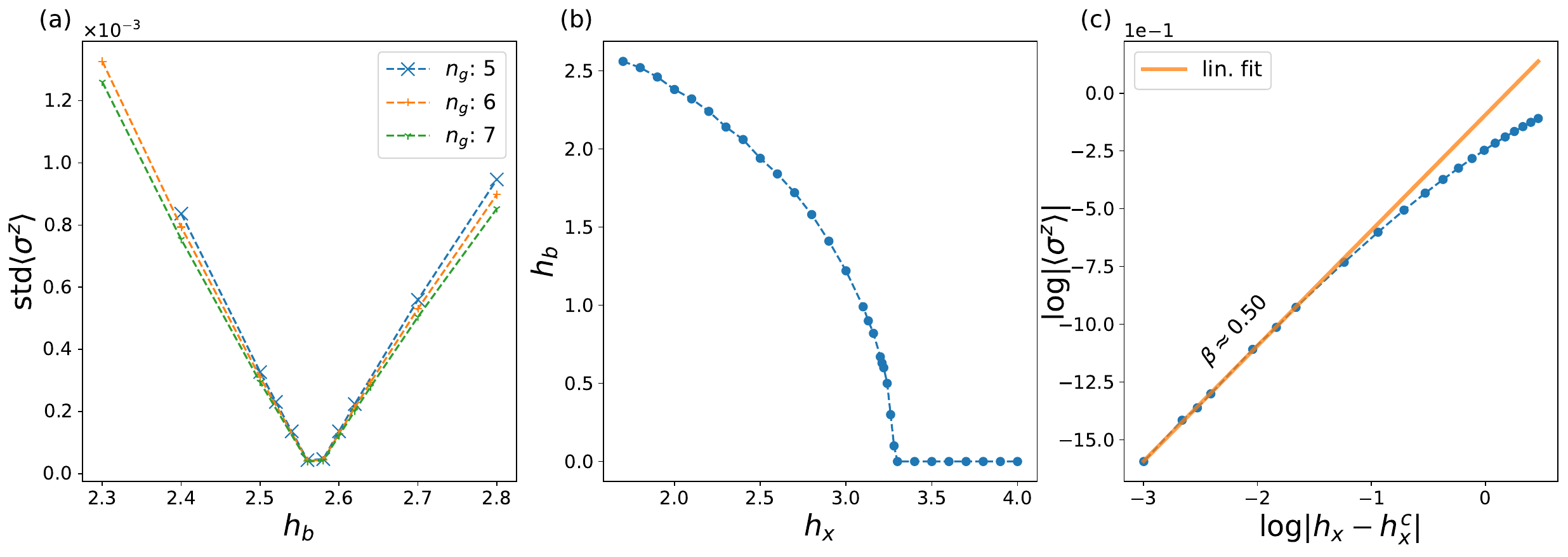}    
        \caption{Data along the effective Bethe Line for $q=4$.(a) Shown the slope of $\mathrm{std}\langle\sigma^z\rangle$ for $h_x = 1.7$ for different generations. (b) Shown the EBL function $h_b(h_x)$. The data are extracted from a tree with $n_g = 5$. (c) Shows the center magnetization in log-log plot and critical exponent $\beta\approx0.5$. The critical point is $h_x^c \approx 2.29$}
        \label{fig:ebl_4}
    \end{figure}
    In Fig.\ref{fig:ebl_4}(c) we show the effective Bethe line for $q=4$

    \newpage
    \section{Connected correlations along the EBL}
    \label{app:cc_EBL}
\begin{figure}[h!]
        \centering
        \includegraphics[width=1\linewidth]{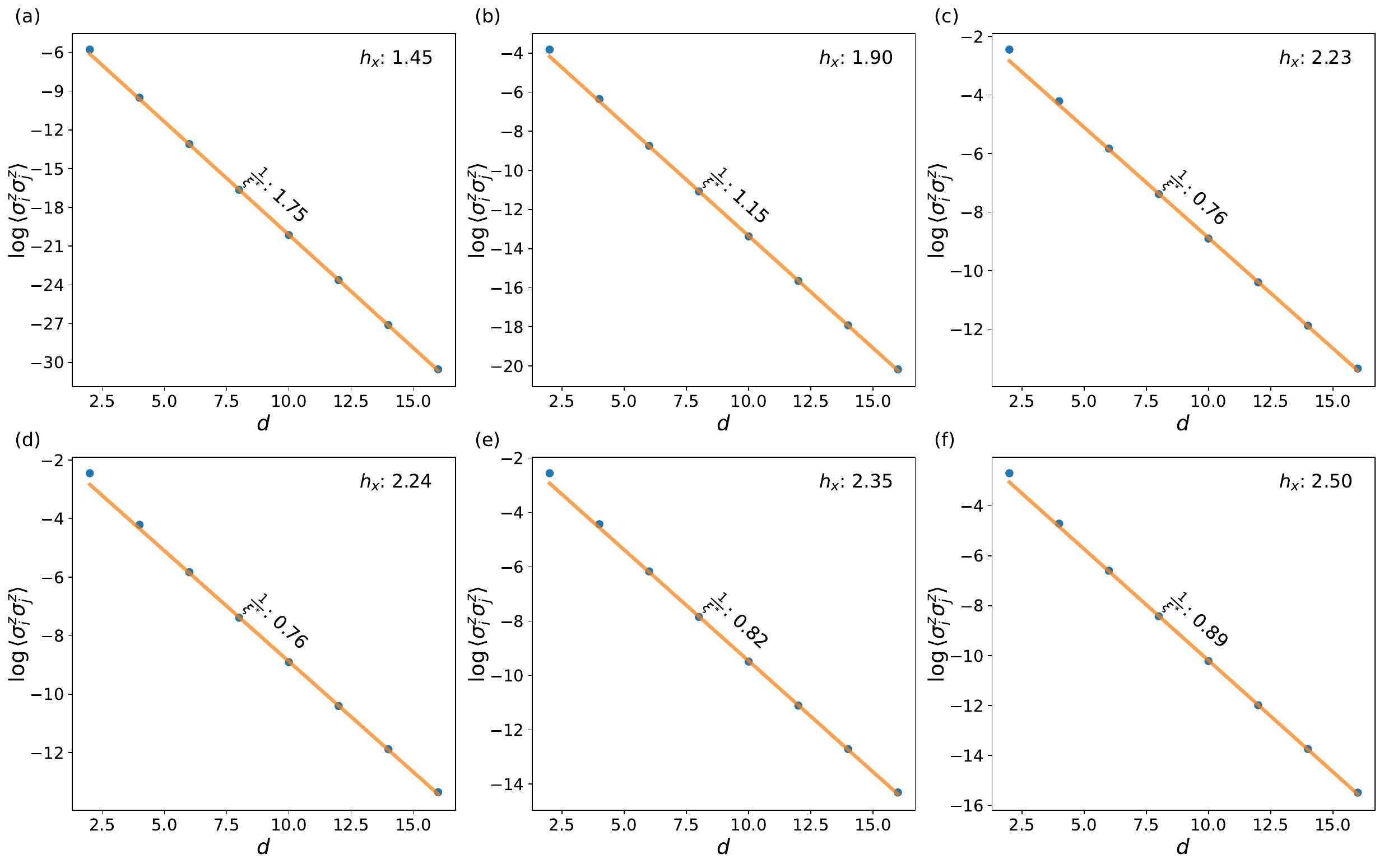}
        \caption{Connected correlations and fit for different points along the EBL in semi-log scale. (a)-(c) are in the ferromagnetic phase. (d)-(f) are in the paramagnetic phase}
        \label{fig:many_cc_ebl}
    \end{figure}

    Along the EBL, the connected correlation functions always decay exponentially. We extract the effective correlation length by fitting the numerical data to Eq.\ref{eq:cc}, excluding the first two points from the fit. In Fig.\ref{fig:many_cc_ebl}, we show representative examples of the connected correlations across the transition at $h_x \approx 2.235$. 

\end{appendix}


\newpage
\clearpage
\bibliography{biblio.bib}


\end{document}